\documentclass[10pt,aps,pra,twoside,twocolumn]{revtex4-2}

\usepackage[english]{babel}
\usepackage{bm}
\usepackage{float}
\usepackage{braket}
\usepackage{amsmath}
\usepackage{mathtools}
\usepackage{mathrsfs}
\usepackage{amssymb}
\usepackage{dcolumn}
\usepackage{fancyhdr}
\usepackage{graphicx}
\usepackage{changepage}
\usepackage[colorlinks=true, allcolors=blue]{hyperref}

\usepackage[dvipsnames]{xcolor}
\newcommand{\white}[1]{{\color{white} #1}}

\usepackage{silence}
\usepackage[most]{tcolorbox}
\newtcolorbox{mytextbox}[1][]{
  sharp corners,
  enhanced,
  colback=white,
  height=0.0cm,
  width=2.0cm,
  #1
}

\makeatletter
\newcommand{\leqnomode}{\tagsleft@true}
\newcommand{\reqnomode}{\tagsleft@false}
\makeatother

\newcommand{\sixj}[6]{\left\{\begin{array}{ccc} #1 & #2 & #3 \\ #4 & #5 & #6 \\ \end{array}\right\}}

\newcommand{\half}{\frac{1}{2}}

\newcommand{\Nmax}{N_{\max}}

\newcommand{\Wigsixj}[6]{\ensuremath{\left\{\begin{matrix}#1 & #2 & #3\cr
#4 & #5 & #6 \end{matrix}\right\}}}

\newcommand{\braketop}[3]{\ensuremath{\left\langle #1 \right| #2 \left| #3 \right\rangle}}

\newcommand{\RedME}[3]{\ensuremath{\langle #1 \| #2 \| #3 \rangle}}

\newcommand{\hw}{\hbar\Omega}

\newcommand{\iu}{\mathrm{i}\mkern1mu}

\begin{document}

\title{Proton and neutron elastic scattering at low energies\\
on He targets from \textit{ab initio} optical potentials 
}
\author{Darin C. Mumma$^{1}$}
\author{Matthew B. Burrows$^{1}$}
\author{Kristina D. Launey$^{1}$}
\author{Daniel Langr$^{2}$}
\author{Tom\'{a}\v{s} Dytrych$^{3}$}
\affiliation{$^{1}$Department of Physics and Astronomy, Louisiana State University, Baton Rouge, LA 70803, USA}
\affiliation{$^{2}$Department of Computer Systems, Faculty of Information Technology, Czech Technical University in Prague, Prague 16000, Czech Republic}
\affiliation{$^{3}$Nuclear Physics Institute of the Czech Academy of Sciences, 250 68 Re\v{z}, Czech Republic}
\date{November 6, 2026}

\begin{abstract}

We construct and discuss \textit{ab initio} nucleon-nucleus optical potentials at low energies for $^{3,4,6}$He targets. In this work, we use the \textit{ab initio} SA-NCSM/GF approach that combines the \textit{ab initio} symmetry-adapted no-core shell model with the Green's function technique to construct optical potentials, and extend this formulation to proton scattering and targets with nonzero spin. We show that these optical potentials reproduce experimental differential cross sections and phase shifts for proton and neutron elastic scattering remarkably well. The \textit{ab initio} SA-NCSM/GF approach provides nonlocal, energy dependent and dispersive optical potentials, suitable for the astrophysically relevant regime of low energies and for exotic nuclei, where experiments are difficult and data is often unavailable.

\end{abstract}

\maketitle

\section{Introduction}

A key ingredient of models for nuclear reactions is the interaction between the reaction fragments \cite{thompson:09book}, also known as clusters. This inter-cluster effective interaction (often called ``optical potential", if dispersive \cite{feshbach:58,feshbach:92,dickhoffbook}) has been historically formulated as a parameterized Woods-Saxon potential that has been typically fitted to elastic-scattering experimental data (see, e.g.,~\cite{koning03}). While this approach has been very successful in describing reactions at higher projectile-energies, isolated low-lying resonances require a special treatment. Furthermore, the need for reliable modeling of nuclear reactions at low energies has becoming increasingly recognized~\cite{JohnsonKDL20}, as one moves away from stability, where uncertainties become uncontrolled since elastic-scattering data does not uniquely constrain the optical potential \cite{LovellN15}. This is especially important for current and planned experiments with rare isotope beams, and for understanding astrophysical processes that proceed through short-lived isotopes at low reaction energies (e.g., see Refs.~\cite{AArcones2017,arnould:20,PDescouvement2020}). 

\vspace{0.1cm} To address this in studies of single-nucleon elastic scattering,  first-principle (\textit{ab initio}) approaches (see Refs.~\cite{JohnsonKDL20,NavratilSQ16, BaccaSP14,Launey_SAreview} for reviews) have been recently employed \cite{NollettPWCH07,HagenDHP07,quaglioni08,ElhatisariLRE15,QuaglioniN09,HupinG2014,mercenneldeqsd21,AFlores2023,KKravvaris2023,KKravvarisPLB2024,KKravvarisPRC2024,burrowslmbsdl24,MAtkinson2025,AFlores2025,GSargsyanPLB2025,MVorabbi2025}. These approaches directly compute phase shifts and cross sections, without the explicit construction of optical potentials, which are in demand and widely used in current few-body reaction codes. Complementary to such developments, one can utilize a many-body approach to reduce the many-body problem of the composite system of clusters to few-body degrees of freedom, by extracting the effective interactions between the clusters. Specifically, recent studies have derived optical potentials, or in general, inter-cluster effective interactions, by utilizing many-body approaches with realistic inter-nucleon interactions, typically derived in the chiral effective-field theory (see, e.g., \cite{BedaqueVKolck02,EntemM03,Epelbaum:2014sza,RevModPhys.92.025004}), without the need to fit interaction parameters in the nuclear medium. For low energies, these models build on pioneering work such as the Green’s function (GF) formulation \cite{CAPUZZICM00} and the dispersive optical model \cite{mahaux:1986zz,mahaux1991,dickhoffbook}. Successful recent applications of the GF method include \textit{ab initio} nucleon-nucleus (NA) potentials for elastic scattering at low projectile energies ($\lesssim 20$ MeV per nucleon) based on the coupled-cluster method \cite{RotureauPD17,RotureauPD18} and the self-consistent Green’s function method \cite{idini19} for closed-shell nuclei, and the symmetry-adapted no-core shell model (SA-NCSM/GF)~\cite{burrowslmbsdl24}. In addition, GF developments for neutron scattering cross sections have been recently applied to a $^{24}$Mg target using microscopic approaches~\cite{JBostrom2025,GSargsyanPRC2025}. The elastic scattering cross sections of Ref.~\cite{GSargsyanPLB2025} have been further augmented through SA-NCSM calculations with a good agreement to experimental data~\cite{Launey_SAreview}.

In this work, we expand the SA-NCSM/GF framework, outlined in Ref.~\cite{burrowslmbsdl24} for neutron elastic scattering off the $^4$He $0^+$ ground state, and construct  \textit{ab initio} nucleon-nucleus optical potentials between a projectile that can be either a proton or neutron ($a=1$), and a target of mass $A$ with any total angular momentum. These potentials are translationally invariant---that is, there is no center-of-mass (c.m.) contribution---and applicable to a broad range of open-shell spherical and deformed nuclei. This is enabled by the SA-NCSM that can presently reach from light to medium-mass nuclei, up through the Ca region \cite{launeymd_arnps21,Burrows_2025}. This approach combines the Green's function technique with the SA-NCSM \cite{launeydd16,launeymd_arnps21}, which accommodates from single-particle features to collective and clustering correlations in nuclei. In addition, an important advantage of the GF technique is that the NA effective potentials include the information about all near reaction channels  through the GF calculations in the composite $(A\pm a)$ systems. In this paper, we use these optical potentials to study elastic scattering. Specifically, we provide \textit{ab initio} proton scattering differential cross sections, as well as cross sections for neutron scattering for several He targets, including the challenging n$+^6$He system.

\section{Theoretical Background}

\subsection{Green's Function Method}

For completeness, we outline the SA-NCSM/GF theoretical framework introduced in Ref.~\cite{burrowslmbsdl24}, focusing on targets with nonzero spin and the presence of the Coulomb interaction for proton projectiles. The SA-NCSM uses SU(3) proper tensors
$a_{(n_\gamma\,0) \ell_\gamma j_\gamma m_\gamma}^\dagger \equiv a_{\gamma m_\gamma}^\dagger$ 
and 
$\tilde a_{(0\,n_\gamma) \ell_\gamma j_\gamma -m_\gamma} =(-1)^{n_\gamma+j_\gamma-m_\gamma} a_{\gamma m_\gamma} $, 
which create and annihilate a particle in a harmonic oscillator (HO) single-particle state $\ket{\gamma m_\gamma}$ labeled by the shell number $n_\gamma$, orbital angular momentum $\ell_\gamma$, total angular momentum $j_\gamma$, and projection $m_\gamma$.  For an $A$-body target ground state with total angular momentum $J_0$, the cluster basis states with total angular momentum $J$ are defined for the ``particle" (+) and ``hole" (-) case as:
\begin{widetext}
\begin{eqnarray}
	\ket{ \Phi^{J^\pi(M)+}_{ J_0\gamma} } \equiv (-1)^{j_\gamma+J_0-J}\left\{ a_{\gamma}^{\dagger} \times \ket{\Psi_{{\rm g.s.},J_0}^A} \right \}^{J^\pi(M)}
	=\sum_{t}\frac{(-1)}{\Pi_{J}}\ket{tJ^\pi(M)}\RedME{tJ^\pi}{a_{\gamma}^{\dagger}}{\Psi_{{\rm g.s.},J_0}^A}, \cr
	\ket{ \Phi^{J^\pi(M)-}_{ J_0 \gamma }} \equiv (-1)^{n_{\gamma}}(-1)^{j_\gamma+J_0-J}\left\{ \tilde a_{\gamma} \times \ket{\Psi_{{\rm g.s.},J_0}^A} \right \}^{J^\pi(M)}
	=\sum_{t}\frac{(-1)^{1+n_\gamma}}{\Pi_{J}}\ket{tJ^\pi(M)}\RedME{tJ^\pi}{\tilde a_{\gamma}}{\Psi_{{\rm g.s.},J_0}^A},
    \label{Eq:pivots}
\end{eqnarray}
where $t$ denotes the complete many-body $A \pm 1$ basis and $\Pi_J = \sqrt{2J+1}$. The total angular momentum projection $M$ is included as a formality, but since results do not depend on it, it is omitted in the labels of later expressions. The target eigenfunctions $\ket{\Psi_{{\rm g.s.},J_0}^A}$ for the ground state (denoted as ``{\rm g.s.}" and often omitted for clarity) are calculated in the SA-NCSM (or any many-body approach), whereas the basis vectors for each $\gamma$ and $J^\pi$, $\ket{ \Phi^{J^\pi \pm}_{ J_0\gamma} }$, are calculated through the single-particle overlaps $\RedME{}{\cdot}{}$ in Eq.~\eqref{Eq:pivots}. For any $J_0$, the Green's function may be written as (see Ref.~\cite{birse:1981ghl}):
\begin{eqnarray}
    G^{J^\pi}_{J_0;\gamma\gamma'}(E)  
    = 
    G^{J^\pi+}_{J_0;\gamma\gamma'}(E)
    +
    (-1)^{2 J_0+1}\sum_{J''}\Pi^2_{J''}
    \sixj{j_\gamma}{J_0}{J''}{j_{\gamma'}}{J_0}{J}
    G^{J''^{\pi}-}_{J_0;\gamma'\gamma}(E),
    \label{GFconfig2nonzeroJ0}
\end{eqnarray}
with
\begin{eqnarray}
	G^{J^\pi+}_{J_0;\gamma\gamma'}(E) \equiv \lim_{\epsilon \rightarrow 0} \bra{\Phi^{J^\pi+}_{ J_0 \gamma }} \frac{1}{E - (H - E^A_0) + \iu\epsilon} \ket{\Phi^{J^\pi+}_{J_0 \gamma' } }
    \,{\rm and}\,\,
   G^{J^\pi-}_{J_0;\gamma'\gamma}(E) \equiv  \lim_{\epsilon \rightarrow 0} \bra{\Phi^{J^\pi-}_{J_0 \gamma' }}  \frac{1}{E - (E^A_0- H) - \iu\epsilon}\ket{\Phi^{J^\pi-}_{J_0 \gamma }},
    \label{GFconfig1}
\end{eqnarray}
\end{widetext}
where $E$ is the incident scattering energy in the c.m. frame relative to the single-nucleon threshold (also denoted here as $E_{\rm c.m.}$ for clarity), $E^A_0$ is the target ground-state energy, $H$ is the Hamiltonian of the ($A \pm 1$)-system, and $\epsilon$ is a parameter introduced in complex analysis to handle the poles near resonance states. Clearly, Eq.~\eqref{GFconfig2nonzeroJ0} for $J_0=0$ coincides with Eq.~(9) of Ref. \cite{burrowslmbsdl24}:\vspace{0.15cm}
\begin{eqnarray}
	\hspace{-0.4cm} G^{J^\pi}_{J_0=0;\gamma\gamma'}(E) &=& 
    G^{J^\pi+}_{J_0=0;\gamma\gamma'}(E) + G^{J^\pi-}_{J_0=0;\gamma'\gamma}(E).
    \label{GFconfig2}
\end{eqnarray}

\vspace{0.15cm} In the SA-NCSM/GF framework, the Green's function is calculated using Eq.~\eqref{GFconfig1}, which has definite total angular momentum $j$ of the projectile. However, in scattering theory, it is more common to first couple the total angular momenta of the clusters, ${J}_0$ and ${I}_{\rm p}$, to a total spin ${I}$, which is then coupled to $\ell$. The Green's function is then expressed in the $I$-scheme as:
\begin{eqnarray}
    &&G^{J^\pi}_{J_0I_{\rm p};In\ell,I'n'\ell'}(E)\cr
    &=&
    \sum_{j j'}
    (-1)^{j-j'}\Pi_{II' j j'}
    \sixj{J_0}{I_{\rm p}}{I}{\ell}{J}{j}\cr
    &\times&\sixj{J_0}{I_{\rm p}}{I'}{\ell'}{J}{j'}
    G^{J^\pi}_{J_0;n (\ell I_{\rm p}) j, n' (\ell' I_{\rm p}) j'}(E),\label{eq:ConfigToCoordGF}
\end{eqnarray} 
with $I_{\rm p}=\frac{1}{2}$ for a single-nucleon projectile. From this, the Green's function can be  constructed in coordinate space following Eq. (10) of Ref.~\cite{burrowslmbsdl24}: $ G^{J^\pi}_{J_0I_{\rm p};I\ell,I'\ell'}(r,r';E) = \sum_{n n'}R_{n \ell}(r) R_{n' \ell'}(r')G^{J^\pi}_{J_0I_{\rm p};In\ell,I'n'\ell'}(E)$, where $r (r')$ designate the inter-cluster distance before (after) scattering. Here, as done in Ref.~\cite{burrowslmbsdl24}, we first invert the Green's function in configuration representation to calculate the inter-cluster nuclear potential, which in the coordinate space is given as:
\begin{widetext}
\begin{eqnarray}
    \hspace{-0.4cm}V_{J_0I_{\rm p};I\ell,I'\ell'}^{J^\pi}(r,r')
    =
    \sum_{n n'}^{n_{\max}}
    \left[\delta_{II'}\delta_{\ell\ell'}\left( 
        E\delta_{nn'}-\bra{n \ell}T_\mathrm{rel}\ket{n'\ell}-\bra{n \ell}V_\mathrm{Coul}\ket{n' \ell} \right) - (G^{-1})_{J_0I_{\rm p};In\ell,I'n'\ell'}^{J^\pi}
    \right]
    R_{n \ell}(r) R_{n' \ell'}(r'), 
    \label{eq:optpot}
\end{eqnarray}
\end{widetext}
where $n_{\max}$ is the highest HO shell available to the $A$ and $A \pm a$ systems (the dependence on $E$ is omitted from the notation). We note that, with $T_{\rm rel}(r)\frac{\delta(r-r')}{r r'}=\sum_{n n'}^\infty R_{n \ell}(r) R_{n' \ell}(r') \bra{n \ell}T_{\rm rel}\ket{n' \ell}$, and similarly for $V_{\rm Coul}(r)$, Eq.~\eqref{eq:optpot} reproduces the equation of motion for the s.p. propagator $V_{\rm Nucl+Coul}(\mathbf{r},\mathbf{r'})=(E-T_{\rm rel}(\mathbf{r}))\delta (\mathbf{r}-\mathbf{r'})-G^{-1}(\mathbf{r},\mathbf{r'})$, which also defines Dyson's equation~\cite{dickhoffbook}. Inverting the Green's function is often challenging, but can be avoided, as done, for example, in approaches that expand in particle-hole ranks to compute the optical potential (self-energy)~\cite{dickhoffbook,WDickoff2004}, or methods that use an iterative procedure in coordinate space (e.g., \cite{GSargsyanPRC2025}). In our framework, we take advantage of the Green's function in configuration representation, which can be calculated exactly within $n_{\max}$, includes all particle-hole excitations, and is a small matrix that is well defined outside the poles or at finite $\epsilon$, and, hence, can be straightforwardly inverted~\cite{burrowslmbsdl24}. 

For a general partitioning $\nu \equiv\left\{(A+a)\alpha;(A)\alpha_0 J_0,(a)\alpha_{\rm p} I_{\rm p}\right\}$ and channel $c \equiv\left\{ \nu I \ell\right\}$ (the labels $\alpha$, $\alpha_0$, and $\alpha_{\rm p}$ denote all other quantum numbers needed to fully characterize their respective states), this potential is then used in the coordinate-space Schr\"{o}dinger equation:
\begin{eqnarray}
    \hspace{-0.5cm}\left[\right.
         -E&+&T_{\rm rel}(r) + V_{\rm Coul}(r)
    \left.\right]
    u_{c}^{J^\pi}(r)\cr
    &+&
    \sum_{c'}\int \mathrm{d}r'r'^2
    V_{cc'}^{J^\pi}(r,r')u_{c'}^{J^\pi}(r')=0,
    \label{eq:TISE}
\end{eqnarray}

\vspace{-0.1cm}\noindent where $u_{c}^{J^\pi}(r)$ is the radial wavefunction of the relative motion for a single channel in units of fm$^{-3/2}$, and for the single-nucleon projectile $c=\{(\alpha_0={\rm g.s.})J_0;I\ell\}$. This Schr\"{o}dinger equation can be solved, e.g., in the ${\bf R}$-matrix approach \cite{DescouvemontB10}, as done in this study. We note that for optical potentials, one uses Eq.~(3.90) in Ref.~\cite{DescouvemontB10}; also, we consider a single partitioning, corresponding to the  quantum numbers of the reaction fragments before the reaction ($\nu = \nu'$), leading to a summation in Eq.~\eqref{eq:TISE} over $I'\ell'$ only.

In this work, the many-body wavefunctions and one-body overlaps are calculated in the SA-NCSM framework. As in typical no-core shell-model approaches, the two parameters that jointly determine the size of coordinate space where the nucleus resides are the maximum number of total HO excitations above the valence-shell configuration included in the model space, $\Nmax$, and the HO energy scale, $\hbar\Omega$. The energy of the $(A\pm a)$-body composite system is calculated across a series of $\Nmax$ values and extrapolated to the infinite model space ($\Nmax\rightarrow\infty$) while $\hbar\Omega$ provides an uncertainty band in a given reaction observable (e.g., cross sections) that is expected to be relatively insensitive to the choice of $\hbar\Omega$ value for converged results. In the SA-NCSM/GF evaluations, we use the ground-state extrapolated energy for the $A$ and $A\pm a$ systems for each $\hw$. As detailed in \cite{burrowslmbsdl24}, using low $\hw$ values in manageable $\Nmax$ model spaces accounts for all the relevant correlations and spatially expanded modes (deformation and clustering), whereas phase shifts and cross sections are in addition very sensitive to the resonance energy. However, at low $\hw$ values, binding energies converge slowly. Hence, using extrapolated energies provides a major advantage, namely, these energies are derived entirely in the theoretical framework without the need for experimental data, while utilizing computational resources only for low $\hw$ values. Most importantly, it is straightforward to use the infinite-space energies in the Green's function approach by simply substituting $E^A_0$ with $E^{A,\infty}_0 - (E^{A\pm1,\infty}_0 - E^{A\pm1}_0)$ in $G^{\pm}$ of Eq.~\eqref{GFconfig1}. This requires extrapolations only for the ground-state energies of the target, $E^{A,\infty}_0$, and the $A\pm 1$ systems, $E^{A\pm 1,\infty}_0$ (Sec.~\ref{sec:modelspace}). According to Eq.~\eqref{GFconfig1}, this procedure preserves the excitation spectrum ($E_{\rm x}$) of each  $A\pm 1$ nucleus. This can be seen if one uses the completeness relation in terms of the eigenvectors of the $A\pm 1$ system with energies $E_k^{A\pm 1}$, $k=0,1,\dots$ \cite{burrowslmbsdl24}: e.g., for the particle case the denominator becomes $E - (E_k^{A+ 1} - E^A_0) + \iu\epsilon=E - (E_{{\rm x},k}^{A+ 1} + E_0^{A+ 1} - E^A_0) + \iu\epsilon$, leading to the infinite-space counterpart $E - (E_{{\rm x},k}^{A+ 1} + E_0^{A+1,\infty} - E^{A,\infty}_0)+ \iu\epsilon$. This corresponds to a constant energy shift of the entire spectrum, while ensuring infinite-space energies are used for the ground states and the single-nucleon threshold.\vspace{-0.25cm}

\subsection{Scattering Theory}

\vspace{-0.2cm}In this section, we briefly outline the steps for calculating differential cross sections, following the discussions in Refs.~\cite{thompson:09book,DescouvemontB10,Blatt1952}, and provide an efficacious algorithm to compute them. For any scattering experiment with two incoming clusters and two outgoing clusters, regardless of partitioning, $u^{ J^\pi}_c(r)$ is the solution to the Schr\"{o}dinger equation according to Eq.~\eqref{eq:TISE}. 
For scattering energy given by the ($A + a$)-body eigenenergy relative to the threshold (a pole of the Green's function), $u^{ J^\pi}_{c=\nu I l}(r)$ coincides with the overlap function:
\begin{eqnarray}
    u_{\nu I \ell}^{ J^{\pi}}(r) 
    = 
    \braket{\Psi_{(A + a)\alpha}^{ J^\pi} | \Phi_{ \nu I \ell} ^{J^{\pi}+}}^*, 
    \label{eq:overlapdef}
\end{eqnarray}

\vspace{-0.1cm}\noindent where $\ket{\Psi_{(A + a)\alpha}^{ J^\pi}}$  are the ($A + a$)-body eigenfunctions  and $\ket{\Phi_{ \nu I \ell}^{J^\pi+}}$ are the cluster basis states  [cf. Eq.~\eqref{Eq:pivots}]. For $E>E_{\nu}$, the exact asymptotic expression for $u^{ J^\pi}_c(r)$ can be written in terms of the incoming ($-$) and outgoing ($+$) spherical Hankel functions $H_\ell^{\pm}(\eta_{\nu},k_{\nu}r)$ with arbitrary coefficient 
$N_c^{J^\pi}$:
\begin{eqnarray}
    ru_{(c_{\rm in})c}^{J^\pi}(r)
    \rightarrow 
    N_{c}^{J^\pi}
    \left[\right.
        \delta_{c_{\rm in}c}H_{\ell}^-(\eta_{\nu},k_{\nu}r)
        -
        S^{J^\pi}_{c_{\rm in}c}H_{\ell}^+(\eta_{\nu},k_{\nu}r)
    \left.\right], \cr
    \label{eq:OverlapAsympt}
\end{eqnarray}

\vspace{-0.1cm}\noindent where $S^{J^\pi}_{c_{\rm in}c}$ is the so-called ${\mathbf S}$ matrix for given $J^{\pi}$ and channel transition $c_{\rm in} \rightarrow c$ for an entrance channel $c_{\rm in}$, and here is calculated directly in the ${\bf R}$-matrix approach~\cite{DescouvemontB10}. In Eq.~\eqref{eq:OverlapAsympt}, $k_{\nu}$ is the wavenumber and $\eta_{\nu}$ is the Sommerfeld parameter for a given reaction energy $E_{\nu}$ in the c.m. frame and reduced mass $\mu_{A,a}=m_Am_a/(m_A+m_a)$ for clusters with masses $m_A$ and $m_a$. Specifically, $k_{\nu} = \sqrt{2\mu_{A,a} E_{\nu}}/\hbar c$ in units of 1/fm (for mass and energy in MeV and using $\hbar c =  197.327$ MeV fm) and $\eta_{\nu} = \mu_{A,a}Z_{0}Z_{\rm p}e^2/(\hbar c)^2k_{\nu}$ calculated as $ \mu_{A,a}Z_{0}Z_{\rm p}\alpha_{\rm f}/(\hbar c) k_{\nu} $, where $\alpha_{\rm f}$ is the fine structure constant.\pagebreak

Without loss of generality, we can accommodate the intrinsic-angle components of the total cluster wavefunction by aligning the incoming beam direction with the $z$-axis. With this choice, $\hat{r}=(\theta,\phi)=\Omega$, and the cluster basis wavefunction
$\Phi_{c}^{J^\pi}(\Omega,\bm{\xi}_0,\bm{\xi}_{\rm p})$ for a channel $c$ is [cf. Eq.~\eqref{Eq:pivots}]:\vspace{-0.12cm} 
\begin{eqnarray}
    &&\Phi_{\nu I \ell}^{J^\pi}(\Omega,\bm{\xi}_0,\bm{\xi}_{\rm p}) \cr 
    &=& 
    \mathcal{A}\left[ 
        \left\{ 
            \Psi_{(A) \alpha_0 J_0}(\bm{\xi}_0)
            \otimes 
            \Psi_{(a) \alpha_{\rm p} I_{\rm p}}(\bm{\xi}_{\rm p})
        \right\}^I 
        \otimes 
        Y_{ \ell } (\Omega) 
    \right]^{J^\pi},\cr
    &&\label{eq:ChannelState}
\end{eqnarray}

\vspace{-0.12cm}\noindent where $\mathcal{A}$ is the antisymmetrization operator which enforces the Pauli exclusion principle and $\Psi_{(A) \alpha_b J_b}(\bm{\xi}_b)$ are the intrinsic states of the target ($b=0$) or projectile ($b=\rm p$) specified by intrinsic coordinates $\bm{\xi}_b$. The expression \eqref{eq:ChannelState} excludes an additional phase of $\iu^\ell$ found in Eq. (2.29) of  Ref.~\cite{DescouvemontB10}, which causes intermediate results to differ, but leads to the same observables. The channel wavefunction may then be written in terms of the cluster wavefunctions and radial wavefunction of the relative motion:\vspace{-0.12cm}
\begin{eqnarray}
    \Psi_{\nu I \ell}^{J^\pi}(\bm{r},\bm{\xi}_0,\bm{\xi}_{\rm p})
    =
    \sum_{\nu' I' \ell'}
    \Phi_{\nu' I' \ell'}^{J^\pi}(\Omega,\bm{\xi}_0,\bm{\xi}_{\rm p})
    \,u_{\nu I \ell,\nu' I' \ell}^{J^\pi}(r).\cr
    \label{eq:PartialWfn}
\end{eqnarray}

\vspace{-0.12cm}\noindent Using the channel wavefunction, the total cluster wavefunction is given in terms of the
Coulomb phase shift $\sigma_{\ell}(\eta_{\nu})=\arg\Gamma(\ell+1+\iu\eta_{\nu})$ by:\vspace{-0.12cm}
\begin{eqnarray}
    \Psi_{\nu M_0 M_{\rm p}}(\bm{r},&&\bm{\xi}_0,\bm{\xi}_{\rm p};E_{0})
    = 
    \iu \frac{\sqrt{\pi}}{k_{\nu}}
    \sum_{J\pi} \sum_{I\ell}\iu^\ell
    \bigl( N_{\nu I \ell}^{J^\pi} \bigr)^{-1}\cr
    \times&&\,
    \Pi_\ell e^{i\sigma_\ell(\eta_{\nu})}
    C_{J_0 M_0 I_{\rm p} M_{\rm p}}^{I M} C_{I M \ell 0}^{J M}\Psi^{J^\pi}_{\nu I \ell}(\bm{r},\bm{\xi}_0,\bm{\xi}_{\rm p}).\cr
    &&\label{eq:TotScattWfn}
\end{eqnarray}

\vspace{-0.12cm}\noindent where projection $M=M_0+M_{\rm p}$. Apart from this construction, the expected asymptotic form for the total cluster wavefunction is:\vspace{-0.12cm}
\begin{eqnarray}
    \Psi_{\nu M_0 M_{\rm p}}(\bm{r},\bm{\xi}_0,&\bm{\xi}_{\rm p}&;E_{\nu})
    \rightarrow
    \Psi_{\rm Coul}(\bm{r})\,
    \Psi_{0 M_0}(\bm{\xi}_0) 
    \Psi_{{\rm p} M_{\rm p}}(\bm{\xi}_{\rm p})\cr
    &+&
    \sum_{\nu'M_0'M_{\rm p}'}
    \frac{e^{\iu(k_{\nu'}r-\eta_{\nu'}\ln2k_{\nu'}r)}}{r}\cr
    &\times&
    f_{\nu' M_0' M_{\rm p}'}^{\nu M_0 M_{\rm p}}(\Omega;E_{\nu})\,
    \Psi_{0' M_0'}(\bm{\xi}_0) 
    \Psi_{{\rm p}' M_{\rm p}'}(\bm{\xi}_{\rm p}),\cr
    &&\label{eq:TotalWfnAsymp}
\end{eqnarray}

\vspace{0.3cm}\noindent where $\Psi_{\rm Coul}(\boldsymbol{r})$ is the total cluster wavefunction for $V_{cc'}^{J^\pi}(r,r')=0$ and $f_{\nu'M_{0}'M_{\rm p}'}^{\nu M_{0} M_{\rm p}}(\Omega;E_{\nu})$ is the scattering amplitude. Comparing Eqs.~\eqref{eq:TotScattWfn} and \eqref{eq:TotalWfnAsymp}, one finds that the scattering amplitude is given by:
\begin{eqnarray}
    &&f_{\nu' M_0' M_{\rm p}'}^{\nu M_0 M_{\rm p}}(\Omega;E_{\nu})\cr
    &=&
    \iu\frac{\sqrt{\pi}}{k_{\nu}}
    \sum_{J \pi}\sum_{I \ell}\sum_{I'\ell'}
    \iu^{\ell-\ell'}\Pi_\ell 
    e^{\iu\left[
        \sigma_\ell(\eta_{\nu})+\sigma_{\ell'}(\eta_{\nu'})
    \right]}\cr
    &\times& 
    C_{J_0 M_0 I_{\rm p} M_{\rm p}}^{I M}
    C_{I M \ell 0}^{J M}
    C_{J_0' M_0' I_{\rm p}' M_{\rm p}'}^{I' M'}
    C_{I' M' \ell' (M-M')}^{J M} \cr
    &\times&
    \left(\delta_{\nu \nu'}
    \delta_{I I'}
    \delta_{\ell \ell'}
    -
    S_{\nu I \ell, \nu' I' \ell'}^{J^\pi}\right)Y_{\ell'}^{M-M'}(\Omega).
    \label{eq:scattamp}
\end{eqnarray}\vspace{-0.1cm}
\white{lorem ipsum}

\noindent Evidently, the scattering amplitude contains dependence on azimuthal angle $\phi$, but if one sums over the (outgoing) $M'_{\rm b}$ labels, this dependence disappears. Averaging over the (incoming) $M_{\rm b}$ labels then gives the unpolarized differential cross section:
\begin{eqnarray}
    \frac{{\rm d}\sigma_{\nu\nu'}}{{\rm d}\Omega}\hspace{-0.15cm}&(&\hspace{-0.15cm}\theta; E_{\nu})
    =
    \frac{1}{\Pi_{J_0 I_{\rm p}}^{2}}
    \sum_{M_0 M_{\rm p}}\sum_{M_0' M_{\rm p}'}\cr
    &\times&
    \left|
        \tilde{f}_{\nu' M_0' M_{\rm p}'}^{\nu M_0 M_{\rm p}}(\Omega;E_{\nu})
        +
        f_{\rm Coul}^{\nu\nu'}(\theta;E_{\nu})\delta_{M_0 M_0'}\delta_{M_{\rm p} M_{\rm p}'}
    \right|^{2},\cr
    &&\label{eq:diffCS}
\end{eqnarray}

\vspace{-0.2cm}\noindent where $\tilde{f}_{\nu' M_0' M_{\rm p}'}^{\nu M_0 M_{\rm p}}(\Omega;E_{\nu})$ uses the flux-corrected S-matrix $\tilde{S}_{\nu I \ell, \nu' I' \ell'}^{J^\pi}=\sqrt{\frac{k_{\nu'}/\mu_{A',a'}}{k_{\nu}/\mu_{A,a}}}S_{\nu I \ell, \nu' I' \ell'}^{J^\pi}$ and $f_{\rm Coul}^{\nu\nu'}(\theta,E_{\nu})$ is the Rutherford scattering amplitude:
\begin{eqnarray}
    f_{\rm C}^{\nu\nu'}(\theta,E_{\nu}) 
    = 
    \frac{-\delta_{\nu\nu'}\eta_{\nu}}{2k_{\nu}\,\sin^{2}(\theta/2)}
    e^{2\iu\left(
        \sigma_{{\ell=0}}(\eta_{\nu})-\eta_{\nu}\ln\sin(\theta/2)
    \right)}.\cr
    \label{eq:rutherford}
\end{eqnarray}

\vspace{-0.1cm}\noindent For neutron scattering ($\eta_{\nu}=0$), Eq.~\eqref{eq:diffCS} reduces
considerably~\cite{Blatt1952}, yielding the Coulomb-free differential cross section:
\begin{eqnarray}
    \frac{d\sigma_{\nu\nu'}}{d\Omega}(\theta;E_{\nu})
    =
    \frac{\pi}{k_{\nu}^2\Pi_{J_0 I_{\rm p}}^2}
    \sum_{L}
    B_L^{\nu\nu'}(E_{\nu}) P_L(\cos\theta),
    \label{eq:nscattformula}
\end{eqnarray}

\begin{figure*}[ht]
\centering
\includegraphics[width=0.329\linewidth]{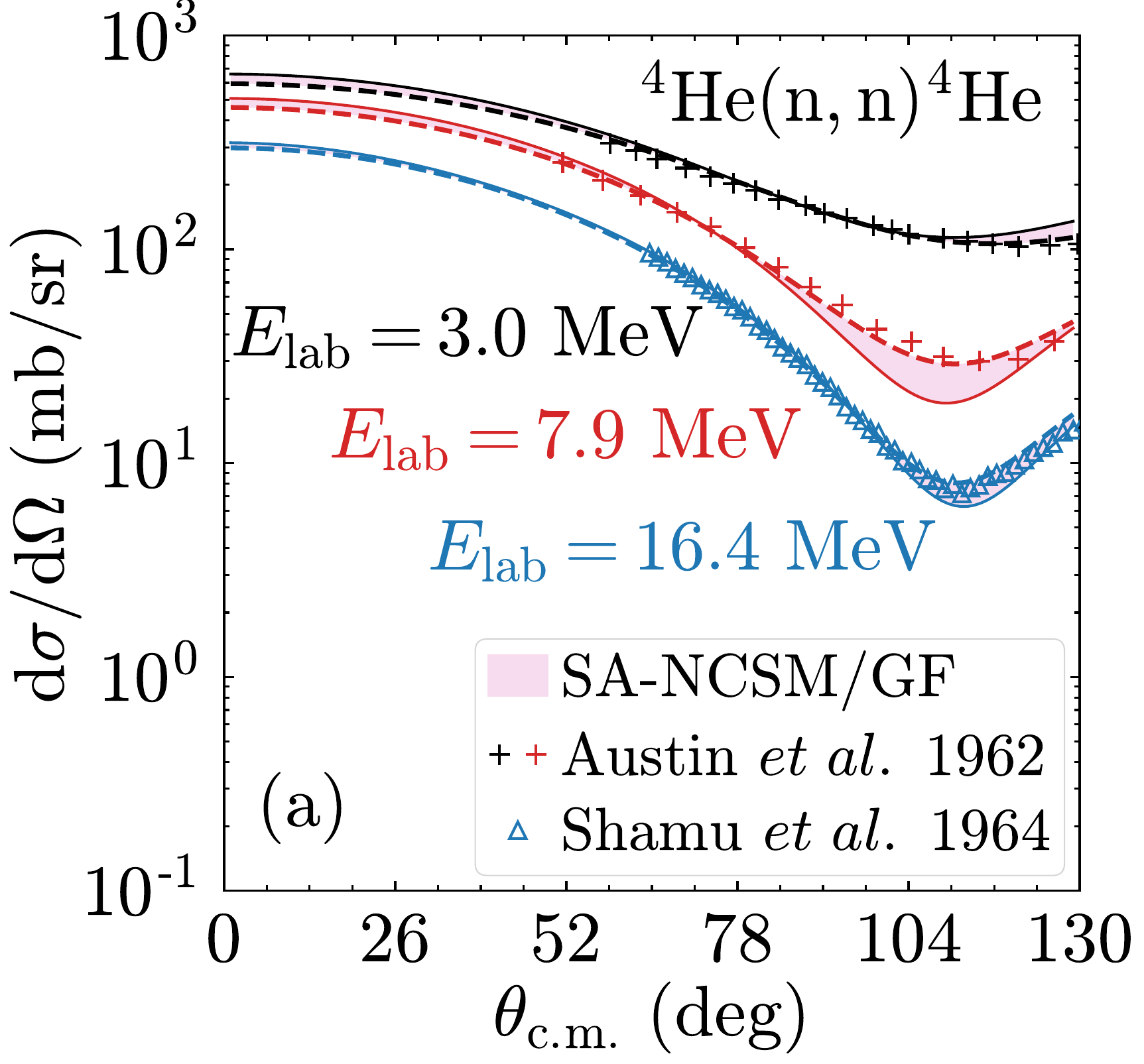}
\includegraphics[width=0.329\linewidth]{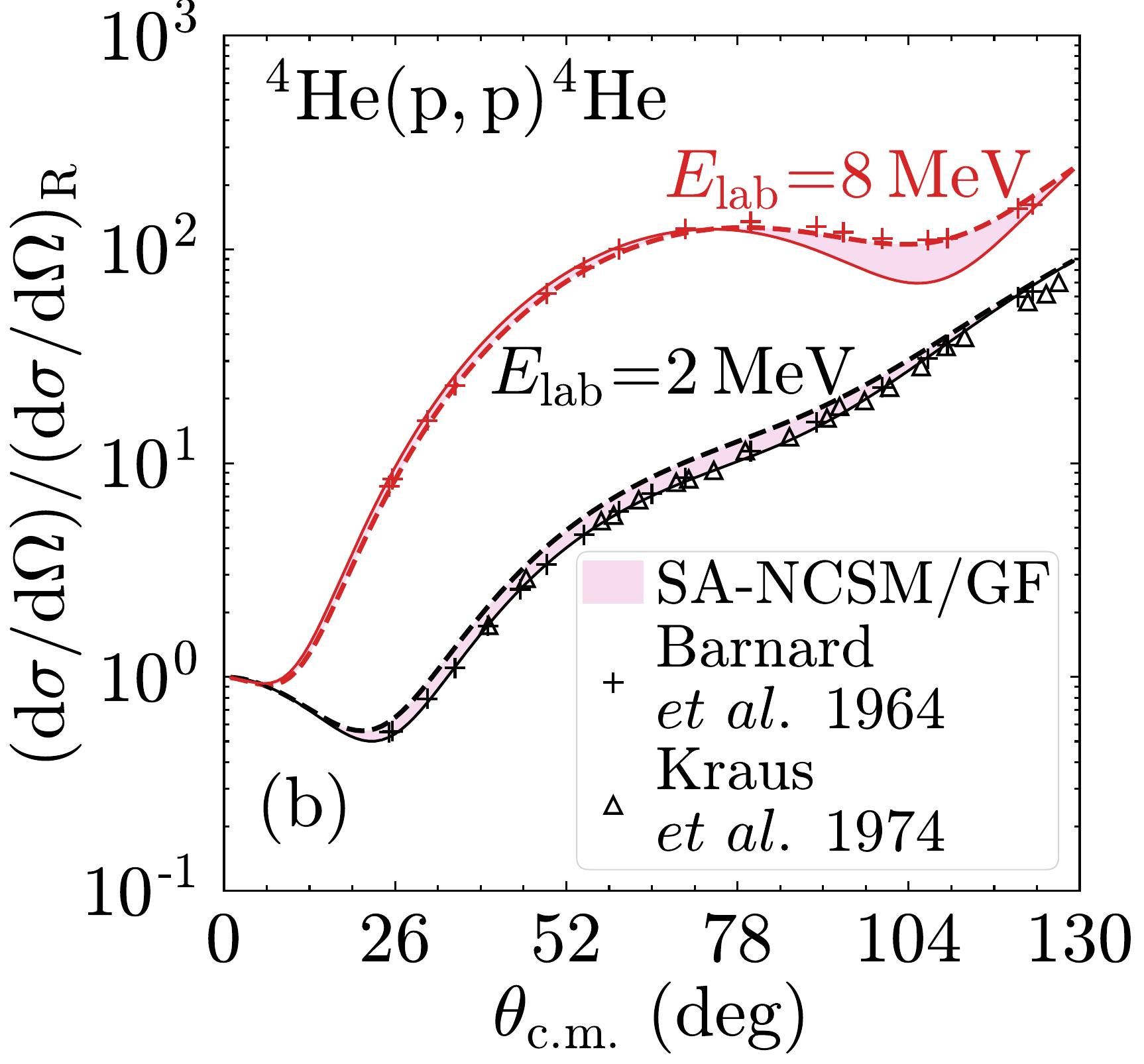}
\includegraphics[width=0.329\linewidth]{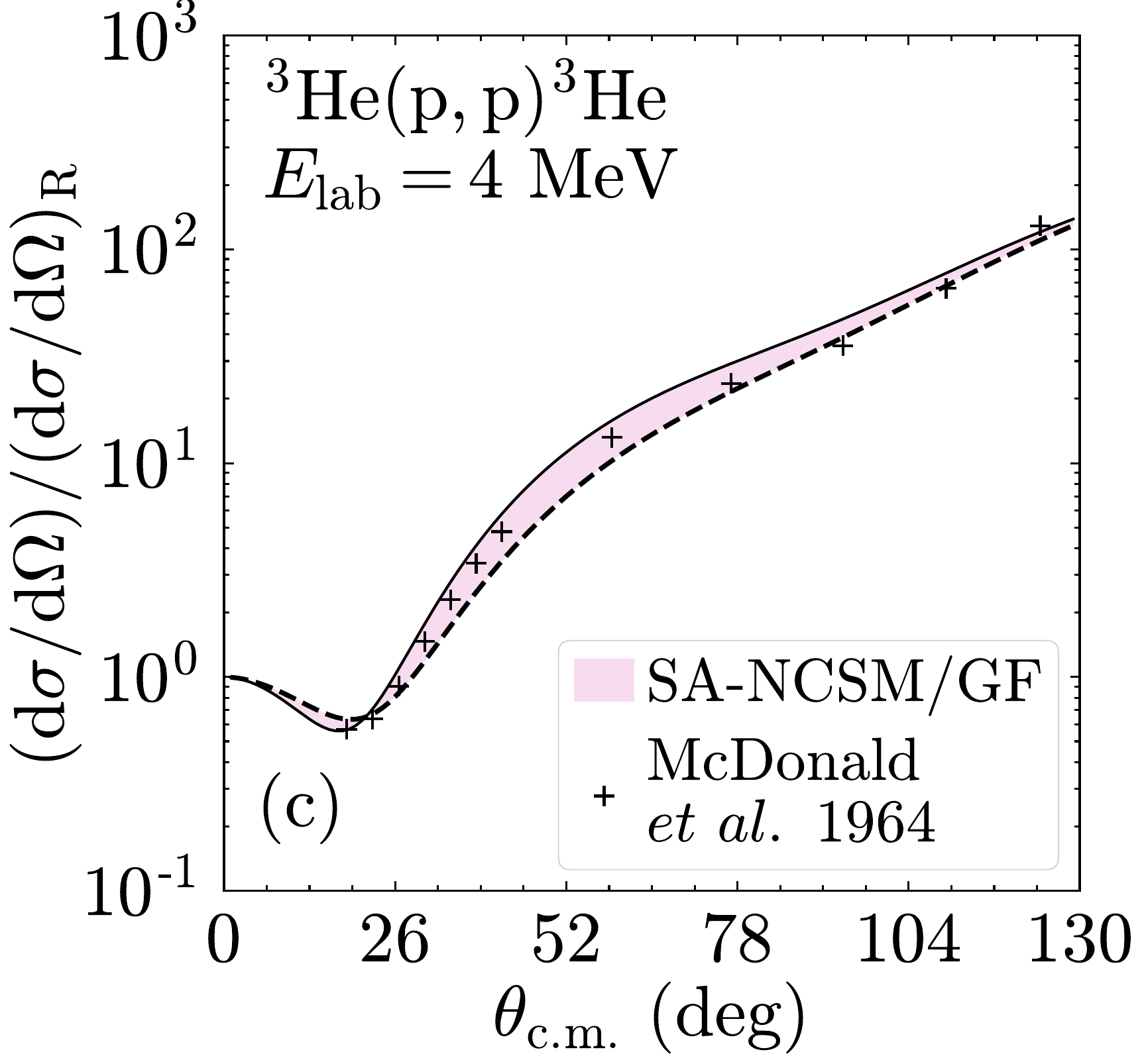}
\vspace*{-0.5cm}\caption{\label{fig:DiffCS_Henuclei_eps0.0_logaxis} Differential cross section vs. projectile scattering angle in the c.m. frame for the reactions (a) $^4\mathrm{He(n,\,n)}^4$He and (b) $^4\mathrm{He(p,\,p)}^4$He (relative to the Rutherford cross section). Calculations include partial waves with $\ell \leq 4$ while using $N_\mathrm{max} = 13$, $\epsilon = 0$ MeV, and $\hbar\Omega = 12{\rm -}20$ MeV, with $\hbar\Omega = 12$ $(20)$ MeV indicated by a dashed (solid) line. (c) Differential cross section, relative to the  Rutherford cross section, for the reaction $^3\mathrm{He(p,\,p)}^3$He. Calculations include partial waves with $\ell \leq 3$ and use $N_\mathrm{max} = 15$, $\epsilon = 0$ MeV, and $\hbar\Omega = 16{\rm -}24$ MeV, with $\hbar\Omega = 16$ $(20)$ MeV indicated by a dashed (solid) line. Experimental results are from Refs.~\cite{Austin1962,Shamu1964,Barnard1964,Kraus1974,McDonald1964}.}
\end{figure*}

\vspace{-0.1cm}\noindent where $P_L$ is the Legendre polynomial determined by coupled orbital momentum $\vec L=\vec \ell+\vec \ell_K=\vec \ell'+\vec \ell_K'$ and $B_L^{\nu\nu'}(E_{\nu})$ is the anisotropy coefficient, given by the expression:
\begin{widetext}
\begin{eqnarray}   
    &B_L^{\nu\nu'}(E_{\nu})
    =\cr
    &\frac{1}{4\pi}
    \sum\limits_{J \pi}\sum\limits_{K \pi_K}
    \sum\limits_{I l \ell_K}
    \hspace{-0.1cm}
    \sum\limits_{\,\,I' \ell' \ell_K'}
    \hspace{-0.1cm}(-1)^{I - I'}\hspace{-0.05cm}
    Z\bigl(\ell  J \ell_K  K, I   L\bigl)
    Z\bigl(\ell' J \ell_K' K, I'  L\bigl)
    \hspace{-0.05cm}
    \left(
        \delta_{\nu\nu'}\delta_{I I'}
        \delta_{\ell \ell'}
        -
        \tilde{S}_{\nu I \ell, \nu' I' \ell'}^{J^\pi}
    \right)
    \hspace{-0.12cm}
    \left(
        \delta_{\nu\nu'}\delta_{I I'}
        \delta_{\ell_K \ell_K'}
        -
        \tilde{S}_{\nu I \ell_K, \nu' I' \ell_K'}^{K \pi_K}
    \right)^\dagger,\cr
    &Z\big(\ell J \ell_K K, I L\big)
    =
    (-1)^{J+K}\Pi_{\ell \ell_K J K}
    C_{\ell 0 \ell_K 0}^{L 0}
    \Wigsixj{\ell}{\ell_K}{L}{K}{J}{I},
    \label{eq:anisocoeff}
\end{eqnarray}
\end{widetext}
where the set of momenta $\left\{K^{\pi_K},\ell_K,\ell'_K\right\}$ are mirrored labels of the set $\left\{J^\pi,\ell,\ell'\right\}$ that appear due to multiplying sums. See Appendix A for an efficient expression of Eq.~\eqref{eq:diffCS} and Appendix B for a connection between Eq.~\eqref{eq:scattamp} and the central+spin-orbit components of the scattering interaction.\pagebreak

\section{Results and Discussions}

\subsection{$^{3}$He and $^{4}$He targets}

We begin by applying the SA-NCSM/GF approach to proton scattering on the spherical target of $^4$He with $J_0=0$. We can thus compare to earlier studies of phase shifts and total cross sections carried forward in alternative approaches, and also to earlier SA-NCSM/GF results for neutron scattering on $^4$He~\cite{burrowslmbsdl24}. Furthermore, we present the first \textit{ab initio} differential cross sections, including those for the $^3$He target with $J_0=\half$.

Specifically, we perform SA-NCSM/GF calculations with NNLO$_{\rm opt}$ \cite{Ekstrom13} in $\Nmax=13$, across a range of $\hbar\Omega = 12{\rm -}20$ MeV. Model uncertainties are reported from $\hw$ variations, which are expected to decrease with larger model spaces and become zero when the exact solution (for the given chiral potential) is reached. From the scattering matrix, we evaluate the phase shifts and the differential cross section for the elastic scattering reaction $^4\mathrm{He(p,\,p)}^4$He using Eq.~\eqref{eq:diffCS}. When comparing the total cross sections to experiment, we use the laboratory kinetic energy of the projectile, $E_\mathrm{lab} = E_{\rm c.m.}(\mu/m_N) = E_{\rm c.m.}(A+1)/A$.

We find that the differential cross sections for $^4\mathrm{He(n,\,n)}^4$He, $^4\mathrm{He(p,\,p)}^4$He, and $^3\mathrm{He(p,\,p)}^3$He calculated in the SA-NCSM/GF are in close agreement with experiment, as shown in Fig.~\ref{fig:DiffCS_Henuclei_eps0.0_logaxis} for projectile laboratory kinetic energies $E_\mathrm{lab}$ from 2 to 16.4 MeV, especially at forward scattering angles. In general, differential cross sections are challenging to model, and the degree of accuracy obtained by the SA-NCSM/GF is remarkable. In addition, at all scattering angles, the spread in the calculated differential cross section arising from the $\hbar \Omega$ variation is very small, even though a significant $\hbar \Omega$ range is considered. We note that this variation for backward angles, while being comparable to that for forward angles, appears to be larger in the figures (Fig.~\ref{fig:DiffCS_Henuclei_eps0.0_logaxis}). The reason is, in the case of $\rm n\,+^4$He, the use of the log scale and the decrease of the cross sections with larger angles. For $\rm p\,+^{3,4}$He, this is expected since we divide by the Rutherford cross section.

\begin{figure}[h]
    \centering
    \includegraphics[width=0.48\textwidth]{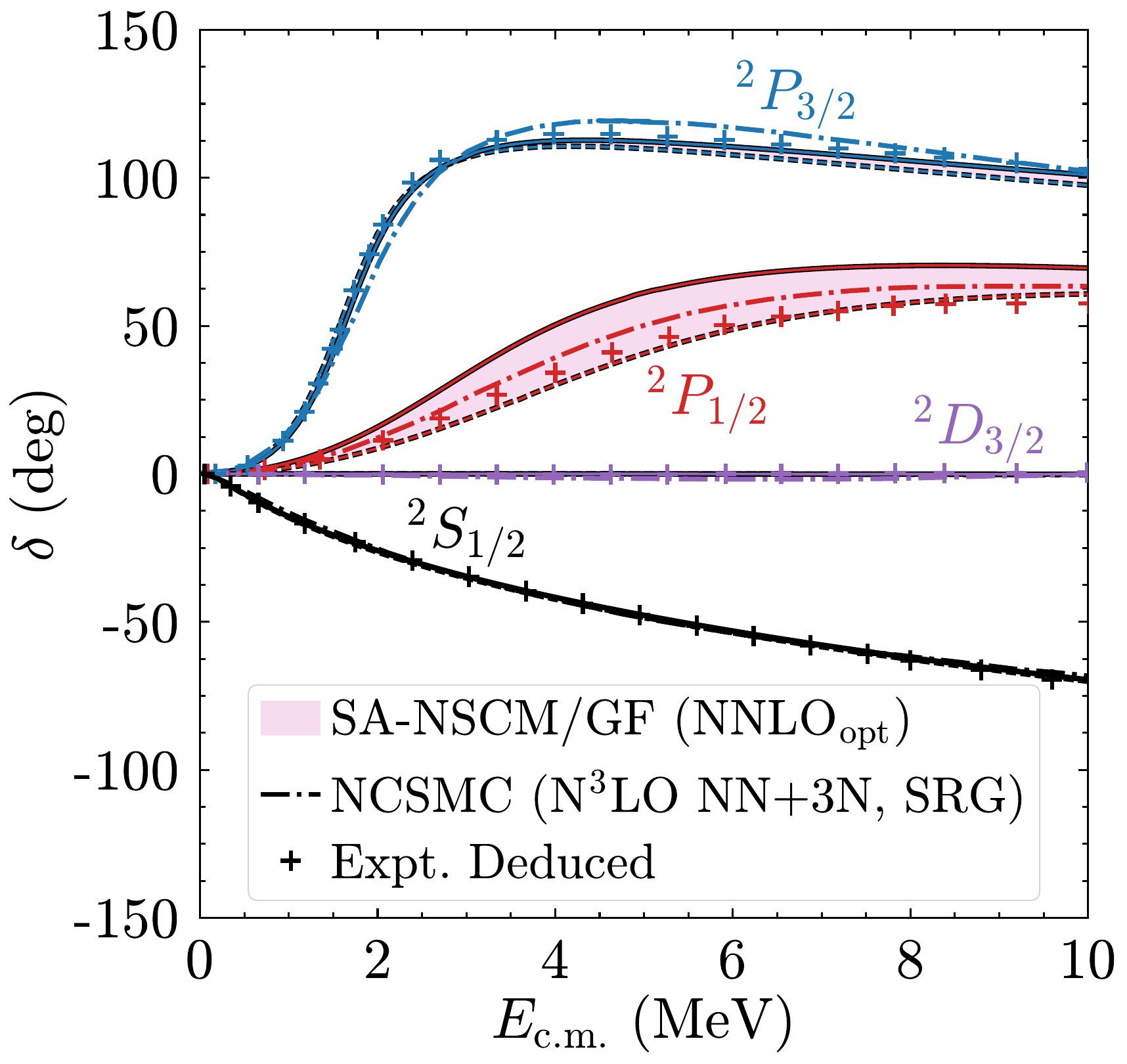}
    \vspace*{-0.5cm}\caption{$^2S_{1/2}$, $^2P_{1/2}$, $^2P_{3/2}$, and $^2D_{3/2}$ phase shifts for the reaction $^4\mathrm{He(p,\,p)}^4$He as a function of the scattering energy in the c.m. frame using the \textit{ab initio} SA-NCSM/GF, compared with the NCSMC theoretical approach ($N_{\rm max}=13$, $\hbar\Omega=20$ MeV) in Ref.~\cite{HupinG2014} and experimentally deduced results \cite{HalePrivComm}. The present calculations use $\Nmax = 13$, $\epsilon = 0$ MeV, and $\hbar\Omega = 12{\rm -}20$ MeV, with $\hbar\Omega = 12$ $(20)$ MeV indicated by a dashed (solid) line.}
    \label{fig:PhaseShifts_p4He_S1_D3_eps0.0}
\end{figure}
Furthermore, the phase shift analyses for $^4\mathrm{He(p,\,p)}^4$He perform remarkably well when compared with both experimentally deduced values and earlier theoretical calculations (Fig. \ref{fig:PhaseShifts_p4He_S1_D3_eps0.0}). The experimentally deduced phase shifts are calculated from experimental total cross sections using an $\mathbf {R}$-matrix evaluation \cite{HalePrivComm}. We find that the $^{2}P_{3/2}$ phase shifts from the SANCSM/GF, e.g., for $\hw = 12$ MeV, yield a threshold energy that agrees with the experimental one to within 70 keV. Our results agree with earlier theoretical calculations carried out in the alternative framework of the NCSMC~\cite{HupinG2014} with a different chiral potential, namely, the SRG-renormalized N3LO-EM NN+3N (cutoff $=$ 2 fm$^{-1}$) \cite{EntemM03,NavratilP2007}, and with those derived in the single-state harmonic-oscillator representation of scattering equations \cite{PhysRevC.98.044624} for different NN interactions \cite{ShirokovMZVW07}. We emphasize that in our calculations, as is usually done in low-energy nuclear simulations, NNLO$_{\rm opt}$ is used without three-nucleon forces, which have been shown to contribute minimally to the three- and four-nucleon binding energies \cite{Ekstrom13}. The agreement with the phase shifts that use N3LO-EM NN+3N further confirms the outcomes of earlier studies, which have shown that the NNLO$_{\rm opt}$ NN potential not only reproduces various experimental nuclear data, but also yields results equivalent to those obtained from chiral potentials that are constructed with significant three-nucleon forces \cite{BakerLBND20,Burrows_2025,sargsyanlbgs2022}.

\subsection{Neutron scattering for $^6$He target}

\begin{figure}[ht]
    \centering
    {\includegraphics[width=0.48\textwidth]{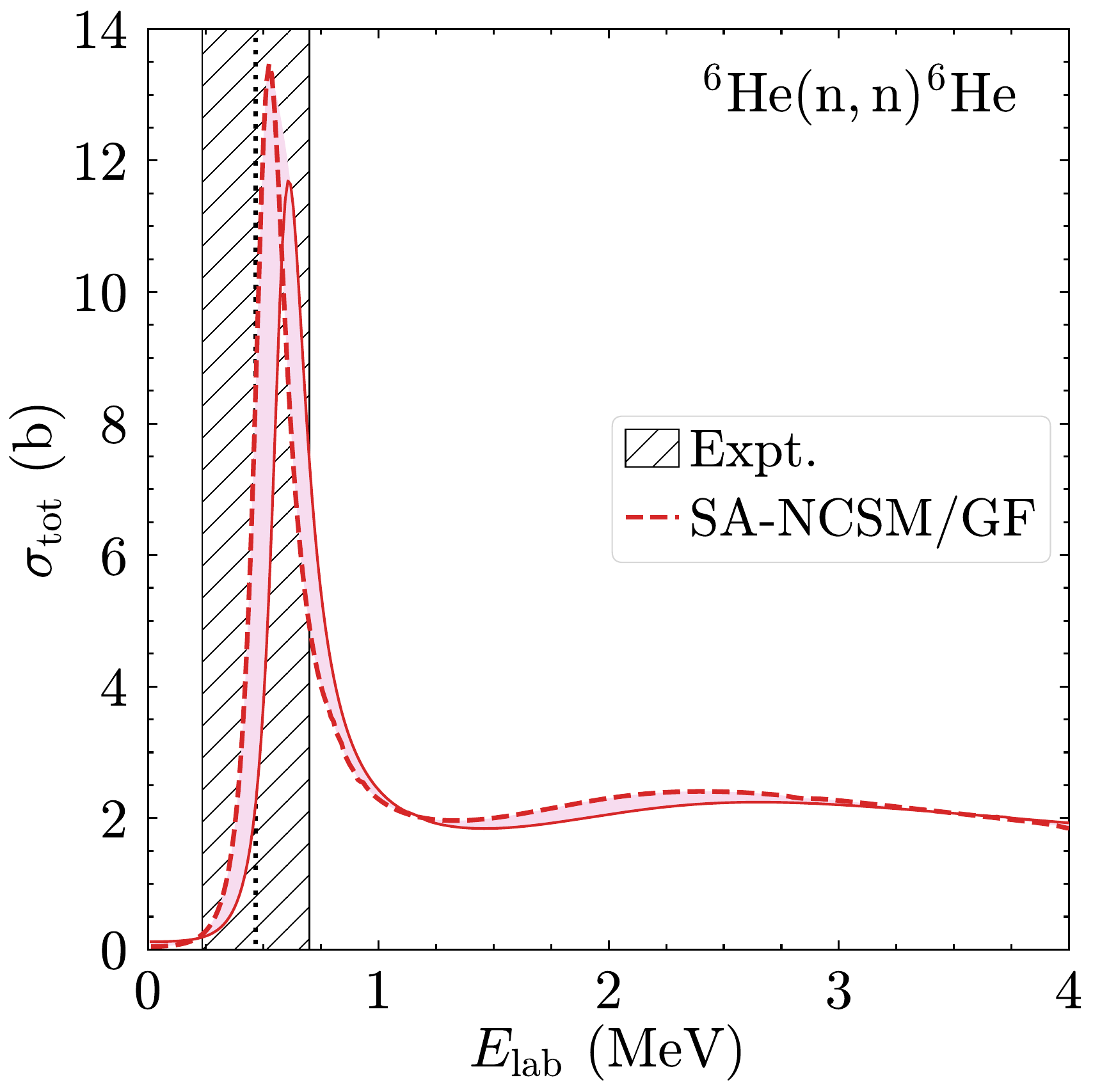}}
    \hspace{-5.61cm}\raisebox{3.25cm}{\includegraphics[width=0.6\linewidth]{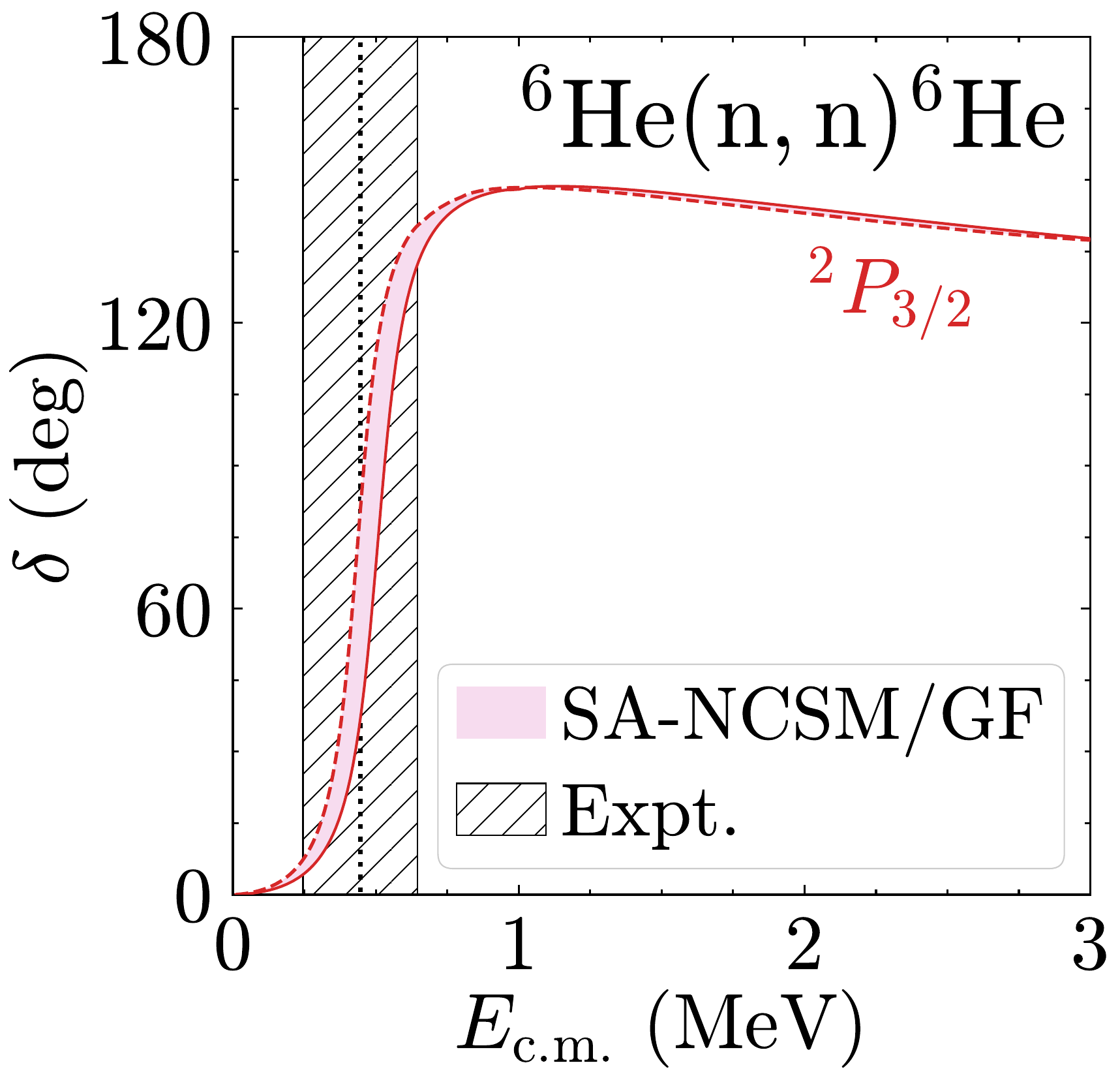}}
    \vspace*{-0.2cm}\caption{Angle-integrated cross section vs. scattering energy in the lab frame for the reaction $^6\mathrm{He(n,\,n)}^6$He, including partial waves $^2S_{1/2}$, $^2P_{1/2}$, $^2P_{3/2}$, and $^2D_{3/2}$. Inset: Phase shift of the dominant partial wave $^2P_{3/2}$ vs. scattering energy in the c.m. frame for the same reaction. Calculations use $N_\mathrm{max} = 9$, $\epsilon = 0$ MeV, and $\hbar\Omega = 12{\rm -}14$ MeV, with $\hw = 12$ $(14)$ MeV indicated by a dashed (solid) line, and the experimental energy and width indicated by the vertical hashed region.}
    \label{fig:CS_n6He_Nmax8_hw12MeV_eps0.0}
\end{figure}

We show that open-shell targets can be studied in the SA-NCSM/GF framework, and we illustrate this for the halo $^6$He nucleus. While halo nuclei are often a challenge for microscopic descriptions, $^6$He has been well described by the SA-NCSM~\cite{launeydd16}. In addition, the ground and first excited state of $^6$He, together with the one-, two-, and three-neutron thresholds in $^7$He, are all within a few MeV of each other. The main challenge is that the $^7$He lowest resonance ($\frac32^-$)  is only $445$ keV above the single neutron threshold. Remarkably, we find that the corresponding peak location in the neutron-$^6$He elastic scattering cross section calculated in the SA-NCSM/GF closely agrees with the $^7$He $\frac32^-$ resonance energy (Fig.~\ref{fig:CS_n6He_Nmax8_hw12MeV_eps0.0}). Specifically, from the SA-NCSM/GF $^2P_{3/2}$ phase shifts (see the inset in Fig.~\ref{fig:CS_n6He_Nmax8_hw12MeV_eps0.0}) we determine a resonance energy of $440$ and $510$ keV for $\hw=12$ and $14$ MeV, respectively. We note that for broader resonances, such as this one, the maximum of the cross section appears at slightly higher energies than the resonance energy~\cite{DescouvemontB10}. In this case, we use extrapolated $^5$He, $^6$He and $^7$He energies from $\Nmax=8$, 10, and 12 calculations based on the Shanks extrapolation method, following Ref.~\cite{burrowslmbsdl24} (see also Sec.~\ref{sec:modelspace}), in the $\hw$ range of $12$-$14$ MeV where cross sections show little dependence on $\hw$, as illustrated in Fig.~\ref{fig:CS_n6He_Nmax8_hw12MeV_eps0.0}. Outside of this $\hw$ range, larger model spaces are desirable to achieve convergence of results and to expand the $\hw$ region where cross sections exhibit little to no $\hw$ dependence. We focus on elastic scattering (real potentials, with imaginary parameter $\epsilon=0$), which is suitable for identifying the energy location of the $^7$He $\frac32^-$ resonance, whereas above ${\sim} 1$ MeV other channels open, requiring in addition imaginary potentials (not included in the figure) and likely the inclusion of $J=5/2$ partial waves. We offer this as an indication of the approach's applicability to open-shell and halo targets.

\subsection{Non-local, energy-dependent optical potentials}

\begin{figure*}[ht]
    \centering
    \includegraphics[width=1.0\linewidth, scale=1.2]{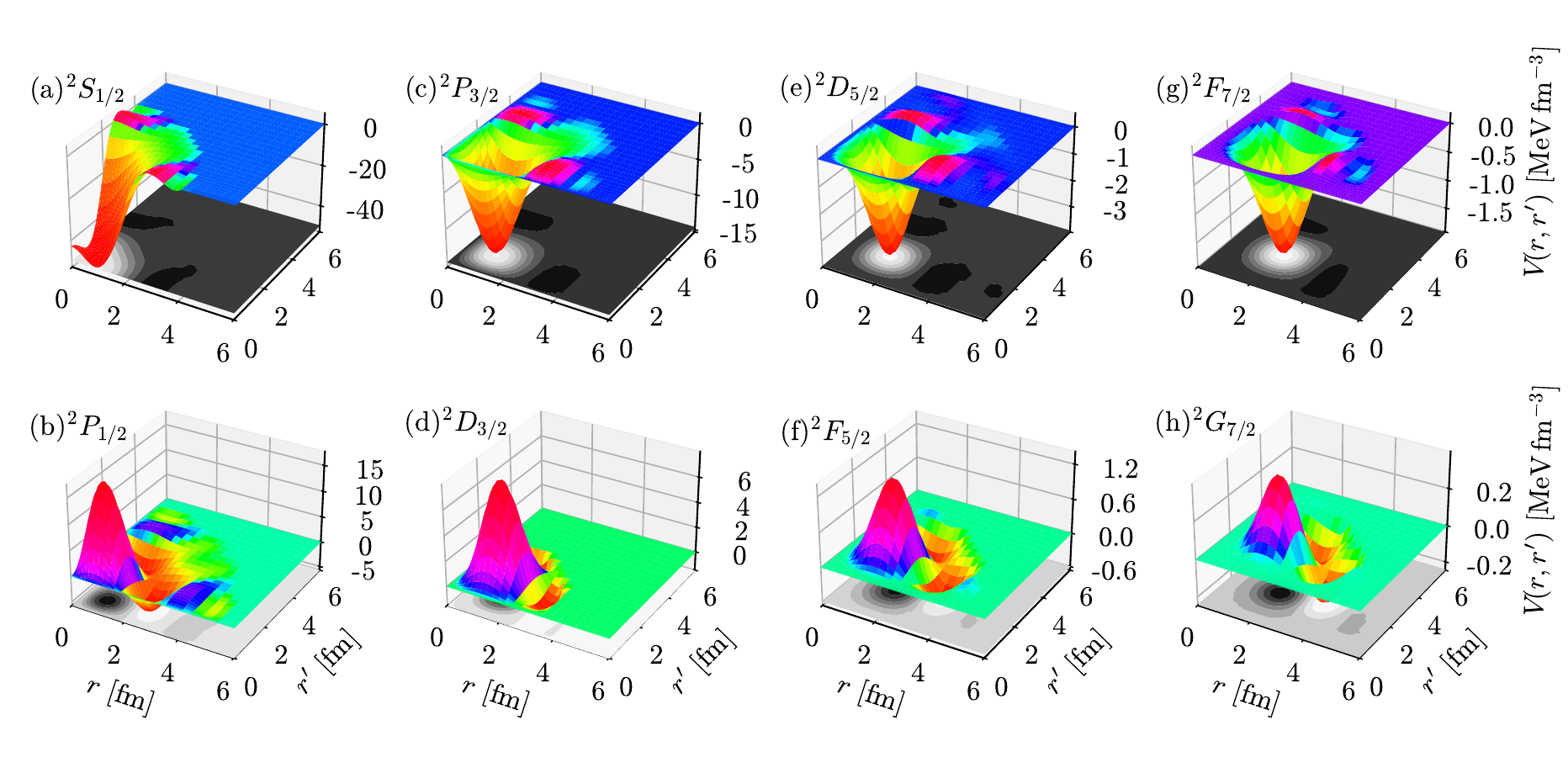}
    \vspace*{-1.0cm}
    \caption{Translationally invariant nonlocal $\rm p\,+^4$He optical potential for the eight $S, P, D, F,$ and $G$ partial waves, calculated in the \textit{ab initio} SA-NCSM/GF for $E_{\rm c.m.}=5.0$ MeV with $\Nmax=13$, $\epsilon=0$ MeV, and $\hbar\Omega=16$ MeV.}
    \label{fig:OP_p4He_E5.0_hw16MeV_eps0.0}
\end{figure*}
The \textit{ab initio} SA-NCSM/GF optical potentials for $\rm p\,+^4$He are highly nonlocal (Fig.~\ref{fig:OP_p4He_E5.0_hw16MeV_eps0.0} for $E_{\rm c.m.} = 5$ MeV). In general, they depend on the scattering energy $E$, however, for this system, we find almost no dependence for  $E_{\rm c.m.}<12$ MeV for all $S_{1/2}$, $P_{1/2}$, $P_{3/2}$, $D_{3/2}$, $D_{5/2}$, $F_{5/2}$, $F_{7/2}$, and $G_{7/2}$ partial waves, except when $E$ is very close to a pole of the Green’s function for $\epsilon$ = 0. Although optical potentials are not observables and cannot be compared exactly between methods and inter-nucleon interactions used, similarities may exist in some features. We find that the $\rm p\,+^4$He optical potentials, which exclude the Coulomb interaction, are nearly identical to the $\rm n\,+^4$He  optical potentials provided in Ref.~\cite{burrowslmbsdl24}. This is consistent with the very small isospin symmetry breaking of the nuclear force.  As such, the proton $S_{1/2}$ partial wave potential (Fig.~\ref{fig:OP_p4He_E5.0_hw16MeV_eps0.0}a) exhibits nonlocal peaks around 2.5 fm, attractive well at smaller distances, and an increase in strength at very small distances, which is similar to the proton $S_{1/2}$ partial wave for another closed-shell target of $^{16}$O when calculated with the NNLO$_{\rm opt}$ and $\hw = 20$ MeV (see Fig. 7 in Ref.~\cite{RotureauPD17}). In addition, the potentials in Fig.~\ref{fig:OP_p4He_E5.0_hw16MeV_eps0.0} should not be directly compared with the orthogonalized nonlocal potentials of the NCSM/RGM for the $\rm p\,+^4$He (ground state), since the latter are effective interactions calculated for each channel. Nevertheless, there is similarity in the shape of the $P_{1/2}$ partial wave from the NCSM/RGM (Fig. 8 of Ref.~\cite{QuaglioniN09}) and the one shown in Fig.~\ref{fig:OP_p4He_E5.0_hw16MeV_eps0.0}b, albeit smaller in magnitude, whereas the SA-NCSM/GF optical potential for the $S_{1/2}$ partial wave is very different from the effective interaction calculated in the NCSM/RGM $\rm p\,+^4$He (ground state). Namely, the attractive well of the GF-based $S_{1/2}$ optical potential allows for an additional state which is bound, as in the $\rm n\,+^4$He case in Ref.~\cite{BurrowsM:2020}. This is a result of the ``particle space" used in NCSM/RGM calculations, compared to ``particle and hole spaces" used in the GF technique. As noted in Refs.~\cite{BurrowsM:2020,Launey_SAreview}, both spaces---while differing in the details of the inter-cluster potential---yield comparable observables, as expected.

\subsection{Convergence of cross sections and energies}

\subsubsection{Partial waves}

\begin{figure}[h]
    \centering
    \includegraphics[width=0.35\textwidth]{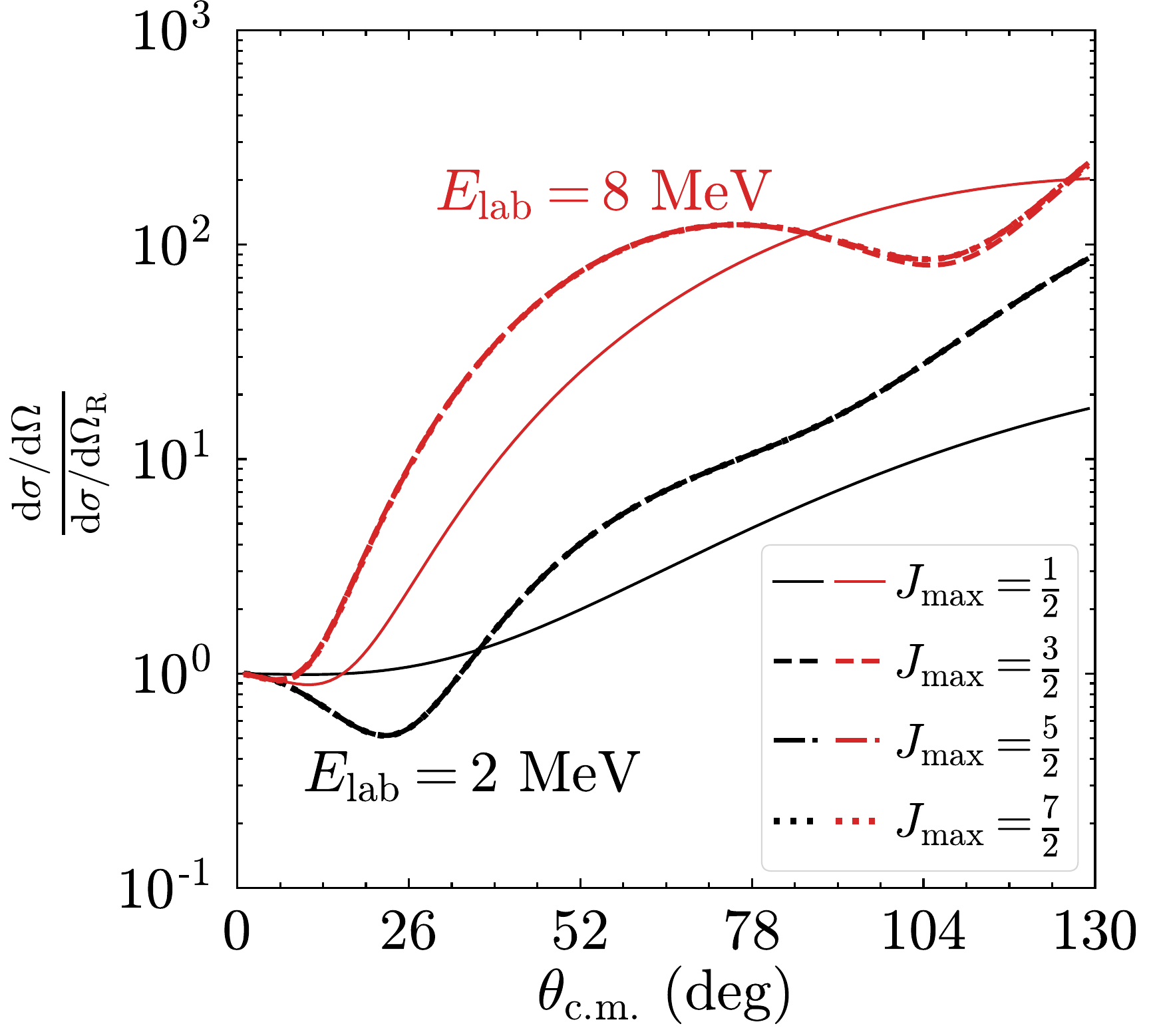}
    \vspace*{-0.2cm}\caption{Differential cross sections for $^4\mathrm{He(p,\,p)}^4$He calculated under the same conditions as those in Fig.~\ref{fig:DiffCS_Henuclei_eps0.0_logaxis}b but truncated to various $J_{\max}$. For example, calculations truncated to $J_{\max} = \frac32$ include the partial waves $^{2}S_{1/2}$, $^{2}P_{1/2}$, $^{2}P_{3/2}$, $^{2}D_{3/2}$ (dashed curves) while calculations truncated to $J_{\max} = \frac72$ also include $^{2}D_{5/2}$, $^{2}F_{5/2}$, $^{2}F_{7/2}$, and $^{2}G_{7/2}$ (dotted curves). The curves for $J_{\max} \geq \frac32$ are practically indistinguishable across the angles.} 
    \label{fig:DiffCS_p4He_E2.0_8.0_hw16_eps0.0_Jtrunc}
\end{figure}

In our cross section calculations, we include partial waves up to a given maximum $\ell_{\max}$ and corresponding $J_{\max}=\ell_{\max}-\half$. To provide \textit{ab initio} predictions, we explore the convergence of cross sections with partial waves, which we illustrate here for the $\rm p\,+^4$He elastic scattering (Fig.~\ref{fig:DiffCS_p4He_E2.0_8.0_hw16_eps0.0_Jtrunc}). We include at most $J_{\max}=\frac72$, but find that only the first few partial waves affect scattering observables as is shown in Fig.~\ref{fig:DiffCS_p4He_E2.0_8.0_hw16_eps0.0_Jtrunc} for the $\rm p\,+^4$He differential cross section. An explanation for this is supplied by Fig.~\ref{fig:OP_p4He_E5.0_hw16MeV_eps0.0}, where we find that the $\rm p\,+^4$He optical potentials decrease by an order of magnitude for each increment to partial wave $J$. In general, we find that for He targets, we need two to three partial waves to achieve convergence of cross sections.

\subsubsection{Model space}
\label{sec:modelspace}

\begin{figure}[ht]
    \centering
    {\includegraphics[width=1.0\columnwidth]{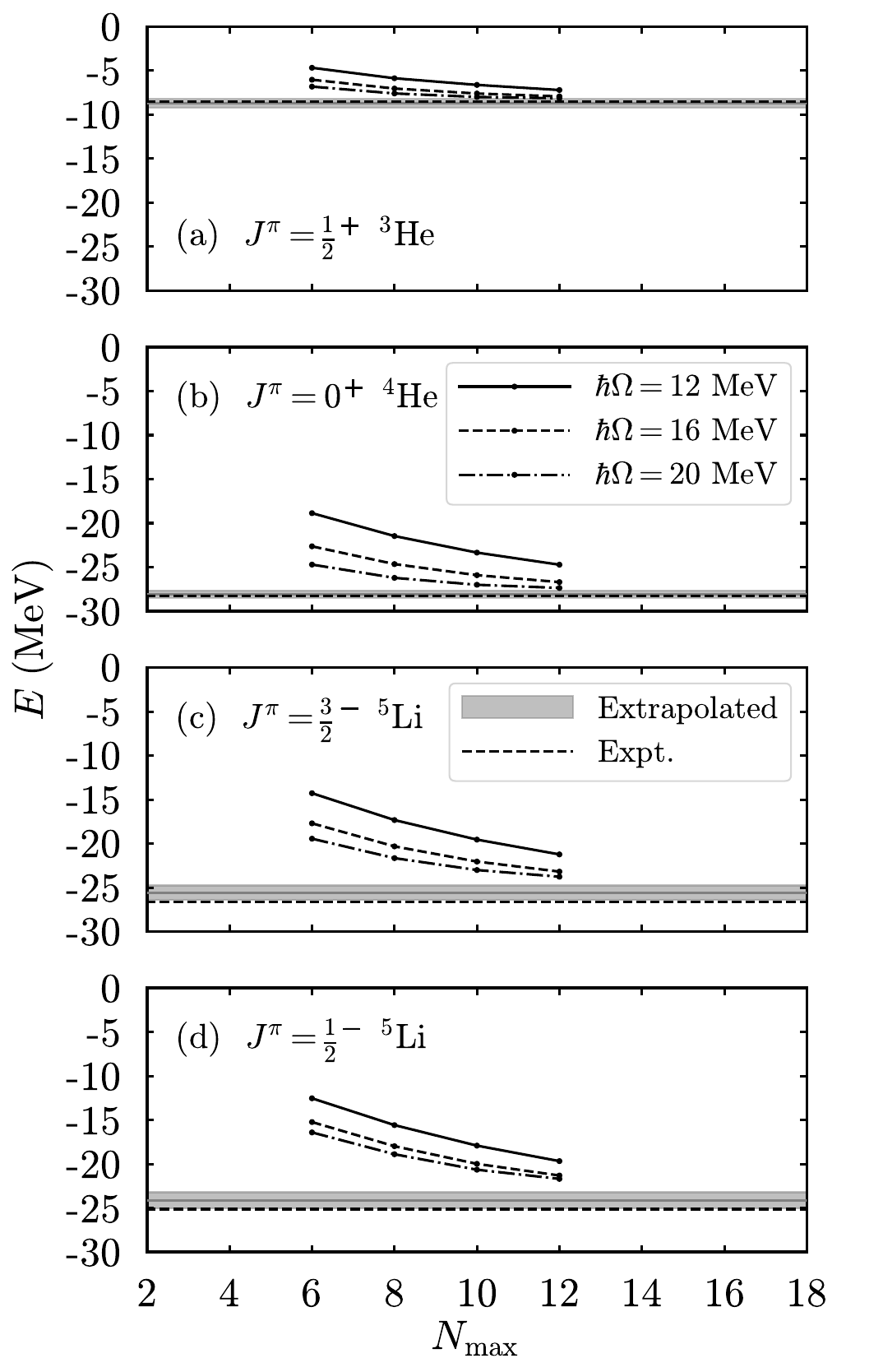}}
    \vspace*{-0.5cm}\caption{Energy of the (a) $\half^+$ ground state of $^3$H, (b) $0^+$ ground state of $^4$He, (c) $^5$Li $\frac32^-$ ground state, and (d) the $^5$Li $\half^-$ resonance with respect to $\Nmax$ across several $\hw$ values (in units of MeV). The extrapolated values across all $\hw$ are given as a band with a centroid in the middle of the band, while the experimental energies are shown as a dashed line.}
    \label{fig:p4He_Nmax_Energy_Dependence}
\end{figure}

Convergence of results with respect to the basis parameters $\Nmax$ and $\hw$ is needed for \textit{ab initio} descriptions in no-core shell-model calculations. We calculate the SA-NCSM energies of $^{2,3}$H, $^{3{\rm -}7}$He, and $^5$Li (see Fig.~\ref{fig:p4He_Nmax_Energy_Dependence} for the $^3$H and $^4$He ground states and the lowest-lying resonances in $^5$Li). The $A\pm1$ ground state and excitation energies are important for the description of all nuclei presented above, as they enter as poles in the Green’s function. We emphasize that, energetically, the poles associated with the $^{2,3}$H ground states lie comparatively far below that of the p$+^{3,4}$He thresholds, that is, by $5.5$ MeV and $19.8$ MeV, respectively. In this case, the poles from the $A-1$ system mainly affect the description of bound states in the optical potential. On the other hand, the pole associated with the $^5$He ground state is only $1.8$ MeV below that of the n$+^6$He threshold, and can have a significant effect on scattering observables. This means that special care must be taken for the $^5$He ground state. Regardless, since the SA-NCSM energies are on a converging trend with respect to $\Nmax$, they can be extrapolated to infinite-space energies in all the nuclei presented here. For this, we use the Shanks extrapolation, which is applied and detailed in Refs.~\cite{Shanks55,dytrychldrwrbb20}. As input, we use calculations with $\Nmax = 11$, 13, and 15 for $\rm p\,+^3$He, $\Nmax = 9$, 11, and 13 for $\rm p/n\,+^4$He and 5, 7, and 9 for $\rm n\,+^6$He. Model uncertainties are estimated across an $\hw = 12{\rm -}20$ MeV range for $^3$He and $^4$He targets.

In Fig.~\ref{fig:p4He_Nmax_Energy_Dependence}, the centroid energies are shown as the midpoint within the $\hw$ region to guide the eye. We note that the convergence of the absolute energies shown in Fig.~\ref{fig:p4He_Nmax_Energy_Dependence} is rapid for $\hw = 16$ and 20 MeV across various model spaces and consistent for the largest model spaces, but is slower for $\hw = 12$ MeV. Nevertheless, the infinite-space extrapolations for all three $\hw$-values agree to within 1 MeV and, as mentioned above, provide reliable predictions of the resonant phase shifts and differential cross sections.

\section{Conclusions}

In this work, we have constructed $\textit{ab initio}$ optical potentials, key ingredients for calculating reaction observables in few-body methods. The optical potentials are obtained in the SA-NCSM/GF framework with chiral potentials, and are illustrated here for several He nuclei. Specifically, we provided the first \textit{ab initio} translationally invariant optical potentials for $\rm p\,+^4$He $\ell \leq 4$ partial waves at low energies. They exceptionally reproduce the experimental phase shifts and differential cross section for $^4\mathrm{He(p, p)}^4$He elastic scattering while also being reasonably independent of $\hbar\Omega$, thereby providing a reliable parameter-free optical potential. We also showed similar agreement in the differential cross sections for $^4\mathrm{He(n, n)}^4$He and $^3\mathrm{He(p, p)}^3$He and in the integrated cross section for $\rm ^6He(n,\,n)^6$He. To construct these potentials, we have extended the SA-NCSM/GF approach to accommodate the projectile-target Coulomb interaction and nonzero total angular momentum of the target ground state. The Green’s function technique uses many-body Hamiltonians with a realistic inter-nucleon interaction, and ensures exact translational invariance, which is critical for light targets. This provides effective nucleon-nucleus potentials that are applicable from stable to unstable nuclei, and contain the information about all near reaction channels, including d and $\alpha$ partitioning. This is achieved through the calculated $A \pm 1$ systems, which, however, makes the problem computationally intensive. In several currently ongoing studies, we explore the use of SA-NCSM selected model spaces to reach heavier nuclei such as $^6$He, $^{12}$C, $^{16}$O, and $^{40}$Ca, as well as applications of the SA-NCSM/GF to deuteron projectiles.

\section*{Acknowledgments}

This work was supported by the U.S. Department of Energy (DE-SC0023532), and in part by the National Nuclear Security Administration through the Center for Excellence in Nuclear Training and University Based Research (CENTAUR) under grant number DE-NA-0004150. This work benefited from high performance computational resources provided by LSU (www.hpc.lsu.edu), the National Energy Research Scientific Computing Center (NERSC), a U.S. Department of Energy Office of Science User Facility at Lawrence Berkeley National Laboratory operated under Contract No. DE-AC02-05CH11231, as well as the Frontera computing project at the Texas Advanced Computing Center, made possible by National Science Foundation award OAC-1818253. We would also like to thank Grigor H. Sargsyan and Alexis Mercenne for useful discussions.\vspace{-0.1cm}

\appendix
\section{Efficient expression for calculating differential cross sections}

\vspace{-0.1cm}Eq.~\eqref{eq:diffCS} is suitable to calculate the differential cross section for any two-body nuclear reaction with arbitrary projectile-target spins. However, as written, it is algorithmically inefficient. In this section, we explain how to achieve speedups with a better expression. We begin by expanding Eq.~\eqref{eq:diffCS}:
\begin{eqnarray}
    &&\frac{d\sigma_{\nu\nu'}}{d\Omega}(\theta;E_{\nu})
    =
    \frac{1}{\Pi_{J_0 I_{\rm p}}^2}
    \sum_{M_0 M_{\rm p}}
    \sum_{M_0' M_{\rm p}'}
    \left|\tilde{f}_{\nu' M_0' M_{\rm p}'}^{\nu M_0 M_{\rm p}}(\Omega;E_{\nu})\right|^2\cr
    &+&
    2\,\mathfrak{Re}\hspace{-0.1cm}\left[
        f_{\rm Coul}^{\nu\nu'*}(\theta;E_{\nu})\frac{1}{\Pi_{J_0 I_{\rm p}}^2}
        \sum_{M_0 M_{\rm p}}
        \tilde{f}_{\nu M_0 M_{\rm p}}^{\nu M_0 M_{\rm p}}(\theta;E_{\nu})
    \right]\cr
    &+&
    \left|f_{\rm Coul}^{\nu \nu'}(\theta;E_{\nu})\right|^2,
    \label{eq:diffCSeqn}
\end{eqnarray}
where the only Coulomb-distorted nuclear scattering amplitude components included are those with $\nu=\nu'$ in the second term since the Coulomb amplitude does not mix partitionings [cf. Eq.~\eqref{eq:rutherford}], which will lead to reductions since $J_0=J_0'$ and $I_{\rm p}=I_{\rm p}'$. The first term is the Coulomb-distorted nuclear cross section, the second term arises from Coulomb-nuclear interference, and the third term is the Rutherford cross section. The sum over spin components can be removed from the first term similarly to the neutron scattering expression given in Eq.~\eqref{eq:nscattformula}. Those in the second term may also be eliminated by invoking two relations over the pairs of Clebsch-Gordan coefficients in Eq.~\eqref{eq:scattamp} as follows:
\begin{eqnarray}
    &&\sum_{M_0 M_{\rm p}}\tilde{f}_{\nu M_0 M_{\rm p}}^{\nu M_0 M_{\rm p}}(\theta;E_{\nu})\cr
    &=&
    \iu\frac{\sqrt{\pi}}{k_{\nu}}
    \sum_{J \pi}
    \sum_{I \ell}
    \sum_{I' \ell'}
    \iu^{\ell-\ell'}\Pi_\ell 
    e^{\iu\left[\sigma_{\ell}(\eta_{\nu})+\sigma_{\ell'}(\eta_{\nu})\right])}\cr
    &\times&
    \left\{\sum_{M_0 M_{\rm p}}C_{J_0 M_0 I_{\rm p} M_{\rm p}}^{I M}C_{J_0 M_0 I_{\rm p} M_{\rm p}}^{I'M'}\right\}\cr
    &\times&
    C_{I M \ell 0}^{J M}C_{I' M' \ell' (M-M')}^{JM}
    \left(
        \delta_{I I'}\delta_{\ell \ell'}-\tilde{S}_{\nu I \ell,\nu I' \ell'}^{J^\pi}
    \right)
    Y_{\ell'}^{M-M'}(\Omega),\cr
    &=&
    \iu\frac{\sqrt{\pi}}{k_{\nu}}
    \sum_{J \pi}
    \sum_{I \ell \ell'}
    \iu^{\ell-\ell'}\frac{\Pi_J^2}{\Pi_{\ell'}}
    e^{\iu\left[\sigma_{\ell}(\eta_{\nu})+\sigma_{\ell'}(\eta_{\nu})\right])}\cr
    &\times&
    \left\{\sum_{M} C_{I M J -M}^{\ell 0}C_{I M J -M}^{\ell' 0}\right\}
    \left(
        \delta_{\ell \ell'}-\tilde{S}_{\nu I \ell,\nu I \ell'}^{J^\pi}
    \right)
    Y_{\ell'}^{0}(\Omega),\cr
    &=&
    \iu\frac{1}{2k_{\nu}}\sum_{J \pi}\sum_{I \ell}\Pi_J^2e^{2\iu\sigma_\ell(\eta_{\nu})}
    \left(
        1-\tilde{S}_{\nu I \ell,\nu I \ell}^{J^\pi}
    \right)
    P_{\ell}(\cos\theta),\cr
    &&\label{eq:noMsumsAllowed}
\end{eqnarray}
where the sums over $M$ may be formally introduced since only terms in which $M=M_0+M_{\rm p}$ are nonzero. This expression now appears in a Legendre sum over orbital angular momentum like Eq.~\eqref{eq:nscattformula}. The Rutherford term in Eq.~\eqref{eq:diffCSeqn} may also be written in this form by inserting $1=\sum_L\delta_{L0}P_L(\cos\theta)$. As a result, one can write an efficient expression for the nuclear differential cross section:
\begin{eqnarray}
    \hspace{-0.1cm}\frac{d\sigma_{\nu\nu'}}{d\Omega}(\theta;E_{\nu})
    =
    \frac{\pi}{k_{\nu}^2\Pi_{J_0 I_{\rm p}}^2}
    \sum_{L}
    \mathscr{B}_L^{\nu\nu'}(\theta;E_{\nu})P_L(\cos\theta),\cr
\label{eq:MEdiffCS}
\end{eqnarray}
where $\mathscr{B}_L^{\nu\nu'}(\theta;E_{\nu}) = B_L^{\nu\nu'}(E_{\nu})+\left(D_L^{\nu\nu'}(E_{\nu})+F_L^{\nu\nu'}(E_{\nu})\right)$ $\sin^{-4}(\theta/2)$, $B_L^{cc'}(E_{\nu})$ is the Coulomb-distorted anisotropy coefficient, $D_L^{\nu\nu'}(E_{\nu})$ is an interference term, and $F_L^{\nu\nu'}(E_{\nu})$ is a Coulomb term. These are each given by:
\begin{widetext}
\begin{eqnarray}
    B_L^{\nu\nu'}(E_{\nu})
    =
    \frac{1}{4\pi}
    \sum_{J \pi}\sum_{K \pi_K}
    \sum_{I \ell \ell_K}
    \sum_{I' \ell' \ell_K'}
    (-1)^{I - I'}
    \hspace{-0.1cm}&e&\hspace{-0.15cm}^{\iu\left[
        \sigma_{\ell}(\eta_{\nu}) 
        +
        \sigma_{\ell'}(\eta_{\nu'})
        -
        \sigma_{\ell_K}(\eta_{\nu})
        -
        \sigma_{\ell_K'}(\eta_{\nu'})
    \right]}
    Z\bigl(\ell  J \ell_K  K, I  L\bigl)
    Z\bigl(\ell' J \ell_K' K, I' L\bigl)\cr
    \times
    \Bigl(
        \delta_{\nu\nu'}\delta_{I I'}
        \delta_{\ell \ell'}
        \hspace{0.1cm}-\hspace{0.1cm}
        \hspace{-0.1cm}&\tilde{S}&\hspace{-0.1cm}_{\nu I \ell, \nu' I' \ell'}^{J^\pi}
    \Bigr)
    \hspace{-0.05cm}
    \Bigl(
        \delta_{\nu\nu'}\delta_{I I'}
        \delta_{\ell_K \ell_K'}
        -
        \tilde{S}_{\nu I \ell_K, \nu' I' \ell_K'}^{K \pi_K}
    \Bigr)^\dagger,\cr\cr\cr
    D_L^{\nu\nu'}(E_{\nu})
    =
    \delta_{\nu\nu'}\frac{\eta_{\nu}}{2\pi}\sum_{J \pi}\sum_I\Pi_J^2
    \Bigl\{
        \Bigl(
            1&-&\mathfrak{Re}\hspace{-0.1cm}\left[S_{\nu I L, \nu I L}^{J^\pi}\right]
        \Bigr)\sin(2\varphi_L(\eta_\nu))
        +
        \mathfrak{Im}\hspace{-0.1cm}\left[S_{\nu I L, \nu I L}^{J^\pi}\right]\cos(2\varphi_L(\eta_{\nu}))
    \Bigr\},\cr\cr\cr
    &F&\hspace{-0.08cm}_L^{\nu\nu'}(E_{\nu})
    =
    \delta_{\nu \nu'}\delta_{L0}
    \frac{\eta_{\nu}^2}{4\pi\Pi_{J_0 I_{\rm p}}^2},   
    \label{eq:3coeffs}
\end{eqnarray}
\end{widetext}
where $\varphi_{L}(\eta_{\nu})=\sigma_{L}(\eta_{\nu})-\sigma_{\ell=0}(\eta_{\nu})+\eta_{\nu}\ln\sin(\theta/2)$ and the $Z$-coefficients are given by Eq.~\eqref{eq:anisocoeff}. The advantage of Eq.~\eqref{eq:MEdiffCS} over Eq.~\eqref{eq:diffCS} is two-fold. First, the elimination of sums over spin components yields remarkable algorithmic speedup, especially when higher spins are involved such as in excited target states or odd-$A$ nuclei. Second, the complete set of symmetries and selection rules are more apparent. Namely, terms which differ by the exchange $\left\{J \ell \ell'\right\}\rightarrow\left\{K \ell_K \ell_K'\right\}$ when $\nu=\nu'$ and $I=I'$ contribute equally to the differential cross section. Additionally, only terms which obey the following five selection rules are nonzero: (i, ii) $\left\{\ell^{(\prime)}+\ell_K^{(\prime)}+L\right\}\bmod2=0$, (iii, iv) $\left|\ell^{(\prime)}-\ell_K^{(\prime)}\right|\leq L\leq \ell^{(\prime)}+\ell_K^{(\prime)}$, and (v) $\left|J-K\right|\leq L\leq J+K$.

\section{Connection between the differential cross section and central (A)/spin-orbit (C) amplitudes}

In Appendix A, we reformat the differential cross section to obtain an efficient expression. While this expression is symmetric, it couples indices and obscures some of the physics. If one instead begins from the scattering amplitude for a limited case, a compact expression can be obtained in terms of functions with clear physical meaning.  

We start by fixing $J_0(')=0$ and $I_{\rm p}(')=\frac{1}{2}$ in Eq.~\eqref{eq:scattamp} (in general, some inelastic channels are still permitted through mass exchange) which yields the identity $C_{0 0 I_{\rm p} M_{\rm p}}^{I_{\rm p} M_{\rm p}}C_{0 0 I_{\rm p} M_{\rm p}'}^{I_{\rm p} M_{\rm p}'}=1$ and the following Coulomb-distorted nuclear scattering amplitude and spin coefficients:\newline
\begin{eqnarray}
    &&\tilde{f}_{\nu' 0 M_{\rm p}'}^{\nu 0 M_{\rm p}}(\Omega;E_{\nu})
    =
    \iu\frac{\sqrt{\pi}}{k_{\nu}}\sum_{\ell}\cr
    &\times&
    \Pi_\ell e^{\iu\left[\sigma_{\ell}(\eta_{\nu})+\sigma_{\ell}(\eta_{\nu'})\right]}
    \beta^{\nu (I_{\rm p}=\frac{1}{2}) \ell M_{\rm p}}_{\nu' (I_{\rm p}=\frac{1}{2}) \ell M_{\rm p}'}(E_{\nu})\,
    Y_{\ell}^{M_{\rm p}-M_{\rm p}'}(\Omega),
    \notag
\end{eqnarray}
with\vspace{-0.1cm}
\begin{eqnarray}
    &&\beta^{\nu (I_{\rm p}=\half) \ell M_{\rm p}}_{\nu' (I_{\rm p}=\half) \ell M_{\rm p}'}(E_{\nu}) 
    = 
    \sum_{J \pi}\cr
    &\times&
    \left\{
        C_{\frac{1}{2}M_{\rm p} \ell 0}^{J M_{\rm p}}
        C_{\frac{1}{2}M_{\rm p}'\ell(M_{\rm p}-M_{\rm p}')}^{J M_{\rm p}}
        \hspace{-0.1cm}
        \left(
            \delta_{\nu\nu'}-\tilde{S}_{\nu (I_{\rm p}=\half) \ell, \nu' (I_{\rm p}=\half) \ell}^{J^\pi}
        \right)
    \right\},\cr
    &&\label{eq:spincoeff}
\end{eqnarray}

\vspace{-0.1cm}\noindent which are invariant to an exchange of $M_{\rm p}(')$ for $-M_{\rm p}(')$. Notably, matrices with elements $\braketop{(I_{\rm p}=\half)M_{\rm p}}{\cdot}{(I_{\rm p}=\half)M_{\rm p}'}$ are size $2\times2$, so the corresponding matrix expressions for the scattering amplitudes are given by:\vspace{-0.1cm}
\begin{eqnarray}
    &&\mathbb{F}^{\nu\nu'}(\Omega;E_{\nu})\cr
    &=&
    \sum_{M_{\rm p} M_{\rm p}'}
    \tilde{f}_{\nu' 0 M_{\rm p}'}^{\nu 0 M_{\rm p}}(\Omega)
    \ket{\half M_{\rm p}}\bra{\half M_{\rm p}'}+\mathbb{F}_{\rm Coul}^{\nu\nu'}(\theta;E_{\nu}),
    \notag
\end{eqnarray}

\vspace{-0.1cm}\noindent and\vspace{-0.1cm}
\begin{eqnarray}
    \mathbb{F}_{\rm Coul}^{\nu\nu'}(\theta;E_{\nu})
    &=&
    \sum_{M_{\rm p} M_{\rm p}'}
    f_{\rm Coul}^{\nu\nu'}(\theta)\delta_{M_{\rm p} M_{\rm p}'}
    \ket{\frac{1}{2}M_{\rm p}}\bra{\half M_{\rm p}'},
    \notag
\end{eqnarray}

\vspace{-0.1cm}\noindent such that the differential cross section may be written:\vspace{-0.1cm}
\begin{eqnarray} 
    &&\frac{d\sigma_{\nu\nu'}}{d\Omega}(\theta;E_{\nu})\cr
    &=&
    \half
    \sum_{M_{\rm p} M_{\rm p}'}\left|f_{\nu' 0 M_{\rm p}'}^{\nu 0 M_{\rm p}}(\Omega;E_{\nu})+f_{\rm Coul}^{\nu\nu'}(\theta;E_{\nu})\delta_{M_{\rm p} M_{\rm p}'}\right|^2\cr
    &=&
    \half
    \left[\left|
        \mathbb{F}^{\nu\nu'}(\Omega;E_{\nu})
    \right|^2\right].
    \label{eq:matrixstuff}
\end{eqnarray}

\vspace{-0.1cm}\noindent Using Eqs.~\eqref{eq:spincoeff} and ~\eqref{eq:matrixstuff}, we now give the total scattering amplitude in matrix form:\vspace{-0.1cm}
\begin{widetext}
\begin{eqnarray}
    \mathbb{F}^{\nu\nu'}(\Omega;E_{\nu})
    &=&
    \iu\frac{\sqrt{\pi}}{k_{\nu}}
    \sum_{\ell}
    \Pi_\ell
    e^{\iu\left[\sigma_\ell(\eta_{\nu})+\sigma_{\ell}(\eta_{\nu'})\right]}
    \begin{pmatrix}
        \beta^{\nu\half\ell\half}_{\nu'\half\ell\half}(E_{\nu})\,Y_\ell^0(\Omega) &
        \beta^{\nu\half\ell\half}_{\nu'\half\ell-\half}(E_{\nu})\,Y_\ell^1(\Omega) &\cr
        \beta^{\nu\half\ell-\half}_{\nu'\half\ell\half}(E_{\nu})\,Y_\ell^{-1}(\Omega) & 
        \beta^{\nu\half\ell-\half}_{\nu'\half\ell-\half}(E_{\nu})\,Y_\ell^0(\Omega) 
    \end{pmatrix}+\mathbb{I}f_{\rm C}^{\nu\nu'}(\theta;E_{\nu}),\cr
    &=&
    \mathbb{I}\left\{
        \frac{\iu}{2k_{\nu}}
        \sum_{\ell}
        \Pi_{\ell}
        e^{\iu\left[\sigma_\ell(\eta_{\nu})+\sigma_{\ell}(\eta_{\nu'})\right]}
        \beta^{\nu\half\ell\half}_{\nu'\half\ell\half}(E_{\nu})\,
        P_\ell(\cos\theta)
        +
        f_{\rm C}^{\nu\nu'}(\theta;E_{\nu})
    \right\}\cr
    &+&
    \mathbb{Y'}(\phi)\left\{
        \frac{-1}{2k_{\nu}}
        \sum_{\ell}
        \frac{2\ell+1}{\sqrt{\ell(\ell+1)}}
        e^{\iu\left[\sigma_\ell(\eta_{\nu})+\sigma_{\ell}(\eta_{\nu'})\right]}
        \beta^{\nu\half\ell\half}_{\nu'\half\ell-\half}(E_{\nu})\,
        P_\ell^1(\cos\theta)
    \right\},\cr
    &=&
    \mathbb{I}A^{\nu\nu'}(\theta;E_{\nu})
    +
    \mathbb{Y'}(\phi)C^{\nu\nu'}(\theta;E_{\nu}),
    \label{eq:AandCscattamp}
\end{eqnarray}

\vspace{-0.1cm}\noindent where we have defined the functions $A^{\nu\nu'}(\theta;E_{\nu})$ and $C^{\nu\nu'}(\theta,E_{\nu})$, $\mathbb{I}$ is the identity matrix, and $\mathbb{Y}'(\phi)$ is a Pauli $y$-matrix rotated about the $z$-axis by an angle $\phi$:\vspace{-0.1cm}
\begin{eqnarray}
    \mathbb{Y'}(\phi)
    =
    \begin{pmatrix}
    0 &
    -\iu e^{\iu\phi} &\cr
    \iu e^{-\iu\phi} & 
    0 
    \end{pmatrix}
    =
    \begin{pmatrix}
    0 &
    \sin\phi-\iu\cos\phi &\cr
    \sin\phi+\iu \cos\phi & 
    0 
    \end{pmatrix}
    =
    \mathbb{X}\sin\phi+\mathbb{Y}\cos\phi.
    \label{eq:Pauliy}
\end{eqnarray}
\end{widetext}

\vspace{-0.1cm}\noindent The meaning of Eq.~\eqref{eq:AandCscattamp} is clear: the scattering amplitude for projectile spin of $\half$ is a (unnormalized) vector on the Bloch sphere with propagation axis along $z$ and polarization axis along $y'=x\sin\theta+y\cos\theta$. This compact form readily yields a convenient expression for the differential cross section:
\begin{eqnarray}
    &&\frac{d\sigma_{\nu\nu'}}{d\Omega}(\theta,E_{\nu})\cr
    &=&
    \half{\rm Tr}\hspace{-0.08cm}\left[\left|
        \mathbb{I}A(\theta;E_{\nu})
        +
        \mathbb{Y'}(\phi)C(\theta;E_{\nu})
    \right|^2\right],\cr
    &=&
    \left|A(\theta;E_{\nu})\right|^2
    +
    \left|C(\theta;E_{\nu})\right|^2\cr
    &\times&
    \half{\rm Tr}\hspace{-0.08cm}\left[
        \mathbb{X}^2\sin^2\phi
        +
        \mathbb{Y}^2\cos^2\phi
        +
        \left\{\mathbb{X},\mathbb{Y}\right\}\sin\phi\cos\phi
    \right],\cr
    &=&
    \left|A(\theta;E_{\nu})\right|^2+\left|C(\theta;E_{\nu})\right|^2,
    \label{eq:AandCdiffCS}
\end{eqnarray}

\vspace{-0.08cm}where we have used Eq.~\eqref{eq:Pauliy} and the fact that Pauli matrices anti-commute with one another, have a square of unity, and are traceless. Returning to the two spin coefficients in Eq.~\eqref{eq:AandCscattamp}, we use Eq.~\eqref{eq:spincoeff} to obtain:
\begin{widetext}
\begin{eqnarray}
    \beta^{\nu\half\ell\half}_{\nu'\half\ell\half}(E_{\nu})
    &=&
    \frac{1}{2\ell+1}\left\{
        (\ell+1)\left(
            \delta_{\nu\nu'}-\tilde{S}_{\nu\half\ell,\nu'\half\ell}^{(\ell+\half)}
        \right)
        +
        \ell\left(
            \delta_{\nu\nu'}-\tilde{S}_{\nu\half\ell,\nu'\half\ell}^{(\ell-\half)}
        \right)
    \right\},\cr
    \beta^{\nu\half\ell\half}_{\nu'\half\ell-\half}(E_{\nu})
    &=&
    \frac{\sqrt{\ell(\ell+1)}}{2\ell+1}\left\{
        \left(
            \delta_{\nu\nu'}-\tilde{S}_{\nu\half\ell,\nu'\half\ell}^{(\ell+\half)}
        \right)
        -
        \left(
            \delta_{\nu\nu'}-\tilde{S}_{\nu\half\ell,\nu'\half\ell}^{(\ell-\half)}
        \right)
    \right\}.
    \label{eq:betacoeff}
\end{eqnarray}
The functions $A^{\nu\nu'}(\theta,E_{\nu})$ and $C^{\nu\nu'}(\theta,E_{\nu})$ can be read directly from Eq.~\eqref{eq:AandCscattamp} in terms of these spin coefficients. Inserting the coefficients yields the expressions: 
\begin{eqnarray}
    A^{\nu\nu'}(\theta;E_{\nu})
    &=&
    \frac{\iu}{2k_{\nu}}
    \sum_\ell
    e^{\iu\left[\sigma_\ell(\eta_{\nu})+\sigma_\ell(\eta_{\nu'})\right]}
    \left\{
        (\ell+1)\left(
            \delta_{\nu\nu'}-\tilde{S}_{\nu\half\ell,\nu'\half\ell}^{(\ell+\half)}
        \right)
        +
        \ell\left(
            \delta_{\nu\nu'}-\tilde{S}_{\nu\half\ell,\nu'\half\ell}^{(\ell-\half)}
        \right)
    \right\}P_\ell(\cos\theta)
    +
    f_{\rm C}^{\nu\nu'}(\theta;E_{\nu}),\cr
    C^{\nu\nu'}(\theta;E_{\nu})
    &=&
    \frac{-1}{2k_{\nu}}
    \sum_\ell
    e^{\iu\left[\sigma_\ell(\eta_{\nu})+\sigma_\ell(\eta_{\nu'})\right]}
    \left\{
        \left(
            \delta_{\nu\nu'}-\tilde{S}_{\nu\half\ell,\nu'\half\ell}^{(\ell+\half)}
        \right)
        -
        \left(
            \delta_{\nu\nu'}-\tilde{S}_{\nu\half\ell,\nu'\half\ell}^{(\ell-\half)}
        \right)
    \right\}P_\ell^1(\cos\theta).
    \label{eq:AandCforJ0of0}
\end{eqnarray}
\end{widetext}
The physical meaning of $C^{\nu\nu'}(\theta;E_{\nu})$ is apparent if one considers a nuclear scattering experiment involving a negligible spin-orbit interaction. In this case, $\tilde{S}^{(\ell+1/2)} \approx \tilde{S}^{(\ell-1/2)}$ and $C^{\nu\nu'}(\theta;E_{\nu})$ vanishes. Hence, $A^{\nu\nu'}(\theta;E_{\nu})$ is the central potential's contribution to the scattering physics while $C^{\nu\nu'}(\theta;E_{\nu})$ is the spin-orbit interaction's contribution.

\bibliography{Hepaper_refs}

\end{document}